\documentclass[superscriptaddress, twocolumn,
amssymb,amsmath,nobibnotes,aps,pre,showkeys,showpacs]{revtex4-1}
 
\usepackage{bm}
\usepackage{latexsym}
\usepackage{amsfonts,amssymb,amsmath} \usepackage{graphicx,epsfig}
\usepackage{psfrag} \usepackage{subfigure} \usepackage{color}
\usepackage{tikz}
\usepackage{verbatim}
\usepackage{ulem} 
\newcommand{\be}{\begin{equation}} \newcommand{\ee}{\end{equation}}
\begin{document}
\title{Floquet topological phase transitions in a kicked Haldane-Chern insulator}

\author{Tridev Mishra}\email{tridev.mishra@pilani.bits-pilani.ac.in}
\author{Anurag Pallaprolu}\email{anurag.pallaprolu@gmail.com}  \author{Tapomoy Guha
  Sarkar}\email{tapomoy1@gmail.com} \author{Jayendra
  N. Bandyopadhyay}\email{jnbandyo@gmail.com} 
\affiliation{Department
  of Physics, Birla Institute of Technology and Science, Pilani
    333031, India.}

\begin{abstract}
We consider a periodically $\delta$-kicked Haldane type Chern insulator with the kicking applied in the 
$\hat{z}$ direction. This is known to behave as an inversion symmetry breaking perturbation, since it 
introduces a time-dependent staggered sub-lattice potential. We study here the effects of such driving on the 
topological phase diagram of the original Haldane model of a Hall effect in the absence of a net magnetic field.
The resultant Floquet band topology is again that of a Chern insulator with the driving parameters, frequency and
amplitude, influencing the inversion breaking mass $M$ of the undriven Haldane model. A family of such, periodically related,
`Semenoff masses' is observed to occur which support a periodic repetition of Haldane like phase diagrams along the inversion 
breaking axis of the phase plots. Out of these it is possible to identify two in-equivalent masses in the 
reduced zone scheme of the Floquet quasienergies, which form the centres of two inequivalent phase diagrams. Further, variation in the 
driving amplitude's magnitude alone is shown to effect the topological properties by linearly shifting the phase diagram of the driven model about the position of 
the undriven case. A phenomenon that allows the study of Floquet topological phase transitions in the system. Finally, we also discuss some issues 
regarding the modifications to Haldane's condition for preventing band overlaps at the Dirac point touchings in the Brillouin zone, in the presence of kicking.
\end{abstract}
\maketitle

\section{Introduction} 
Topology and notions intrinsic to it were introduced into the band
theory of solids through the work of Thouless, Halperin and others
\cite{TKNN,Halperinedge,AvronSeilerPRL83,NiuThoulessPRB85,HatsugaiPRL93}
while theoretically exploring the remarkable phenomenon of the Integer Quantum
Hall Effect (IQHE) \cite{QHE,Laughlingauge}. Many such exotic features were predicted
and identified for other associated phenomena, which went beyond
conventional time-reversal symmetry breaking, such as the Quantum Spin
Hall Effect (QSHE), in graphene and other topological materials
\cite{BernevigZhangPRL06,BernevigHughesZhangSci06,KaneMelePRL05,KaneMeleZ2PRL05}.Experimentally,
this has sparked off a flurry of activity directed towards the
synthesis of materials and nano structures which exhibit such novel
features
\cite{FuKaneMelePRL07,MooreNat10,ChenScirep09,HasanArxiv14,HasanKaneRMP10,QiZhangRMP11}. As
such shaping the field of `Topological Insulators'.  From a
theoretical perspective, the broad objective has been to achieve a
comprehensive classification scheme for these insulators
\cite{AltlandZirnbauerPRB97,SchnyderFurusakiPRB08,KitaevAIP09}.

Graphene, beyond its much touted mechanical and transport properties
\cite{NetoGuineaRMP09,DasSarmaAdamRMP11,GoerbigRMP11}, has shown
itself to be rich in topological features \cite{DelplaceMontambauxPRB11,HatsugaiFukuiPRB06} and various topological aspects of the honeycomb lattice have been
investigated in cold-atom and photonic-crystal setups
\cite{ZhangNat05,LimGoerbigPRA12,JotzuEsslingerNat12,RechtsmanPlotnikPRL13,coldatomHaldane}. Studies of graphene irradiated or periodically driven by circularly
polarized light have revealed rich topological textures beyond those
seen in the undriven case
\cite{OkaAokiPRB09,KitagawaOkaBrataasPRB11,GuFertigArovasPRL11,SuarezTorresPRB2012,IadecolaCampbellPRL13,DelplaceLeonPlateroPRB13,PiskunowUsajPRB14,UsajPiskunowPRB14,PiskunowTorresPRA15,SentefClaassenNat15}. An
entire sub-domain of ``Floquet Topological
Insulators''\cite{CayssolDoraPSS13} has emerged as a result, which constitutes those systems that exhibit topological ground states and
edge phenomena in the presence of certain time-dependent periodic potentials\cite{BukovPolkovnikovAdvPhys14}. They have
offered unprecedented control and freedom to engineer new topological
phases and edge state(in some cases Majorana modes) behaviors
\cite{KatanPodolskyPRL13,WangSteinbergSci13,TongAnGongPRB13,GrushinNeupertPRL14,HeZhangPLA14,ZhouWangGongEPJB14,AnisimovasEckardtPRB15,BenitoPlateroPcaE15,FarrellBarneaPRB16,LongwenChongGongPRB16,XiongGongAnPRB16,KushPRB16,InoueTanakaPRL10,DoraCayssolPRL12,LindnerBergmanPRB13,KunduFertigPRL14,HoGongPRL12,DehghaniOkaPRB14,LagoAtalaTorresPRA15,TitumBergPRX16}
as well as a knob to study topological phase transitions in cold-atom
or photonic crystal setups
\cite{RechtsmanZeunerPoltnikNat13,ZhengZhaiPRA14,ReichlMuellerPRA14,YanLiYangNat15,VerdenyMintertPRA15,LeykamRechtsmanPRL16,RaciunasEckardtPRA16}.
The theoretical classification of Floquet topological insulators and
the identification of valid topological invariants that correctly
characterize the bulk-edge correspondence for these systems is an
on-going effort
\cite{KitagawaBergRudnerPRB10,LeonPlateroPRL13,RudnerLindnerBergPRX13,CarpentierDelplacePRL15,NathanRudnerNJP15,FulgaMaksymenkoPRB16,FruchartPRB16,RoyHarper.arXiv2016}.

Of late, the use of delta-function kicks has also been shown to impart interesting topological
properties in the form of new Floquet topological phases such as
semi-metallic phases in Harper models \cite{BomantaraZhouGongPRE16},
chiral edge modes in Quantum Hall systems \cite{SatijaZhaoPRL14},
appearance of unexpected topological equivalence between spectrally
distinct Hamiltonians \cite{WangHoGongPRE13} as well as generation of
Majorana end modes in 1-D systems
\cite{ThakurathiPatelDiptimanPRB13}. This has led to interest in
studying Dirac systems especially graphene, its nano-ribbons and other
hexagonal lattice models such as the Kitaev model under periodic
driving or
kicking \cite{BabajanovEggerEPJB14,Bhattacharya2016,AgarwalaDiptimanPRB16}.  In fact a very recent work
has studied the effects of periodic kicking on the topological properties of the Qi-Wu-Zhang (QWZ) Chern insulator \cite{WangLiKickChernPRB17} and shown
the emergence of higher Chern number phases in certain cases. In this work 
we shall be considering a form of kicking which is found to introduce a Semenoff like mass, hence no topological nontivialities (in the 
absence of time reversal symmetry breaking) 
in the spectrum of planar Graphene but, shows some promise as far as manipulating the topology of
Haldane-like Chern insulators is concerned. 

The recent success in realizing the Haldane model experimentally \cite{coldatomHaldane} within the framework of ultracold atoms in optical lattices  has opened a doorway to engineering various kinds of Chern insulators, 
 using the paradigm of shaken optical lattices and the Floquet formalism,  and studying topological transitions in them \cite{VerdenyMintertPRA15,RaciunasEckardtPRA16,PlekhanovRouxPRB17}. These realizations offer  an appreciable degree of tunability
and provide an encouraging platform for the study of Haldane systems under periodic driving. We consider these setups as possible avenues for realizing the kind of delta-kicked Haldane 
model which is the centerpiece of our study.  Beyond the cold atom
 setups, an interesting recent experiment \cite{SentefClaassenNat15} drives graphene itself using ultra-fast, short-duration low frequency laser pulses of circularly polarized
 light which open local gaps in the Floquet quasi-energies of the irradiated graphene. This procedure hints at the creation of local Haldane like band structures
 but their topological classification has issues that need to be addressed. A more viable candidate for an actual material realization of 
 the Haldane model is presented in \cite{WrightNatSRep13} where, a honeycomb lattice, specifically Silicene, with out-of-plane staggering
 of sublattice sites, effectively realizes Haldane's prescription of a staggered magnetic field upon being subjected to an in-plane magnetic field which could
 be made very weak. Other interesting proposals exist, that realize Chern insulators either using electron correlations at low dimensions, in say double perovskite
 hetero-structures \cite{CookParamekanti14}, or the notion of in-plane magnetic fields, such
 as in perovskite monolayers \cite{CookPRB16} and laterally patterned $p$-type semiconductor hetero-structures with low-symmetry interfaces \cite{LiSushkovPRB16}.
 The proposal in \cite{WrightNatSRep13}, along with the suggestions in \cite{AgarwalaDiptimanPRB16} that outline methods to implement a kicking using hexagonal boron nitride over
  graphene \cite{ShafiqAdamNat15,OrtixYangPRB12,WeinbergStaarmann2DMat16}, provide the broad experimental context in which our system has some hope of being realized. This motivates our academic interest in
 the study undertaken here. 

 In this paper, we begin with an overview of various features of the Haldane model, broadly describing its spectral and topological aspects in Sec.\ref{Halmod}.
 This is followed by the description and analysis of our choice of a kicked Haldane model in Sec.\ref{kickhalmod} and a brief introduction to the computation
 of the Chern topological invariant in Sec.\ref{cherninv}. A detailed exposition of the various properties and behavior of the kicked model is provided in the
 Sec.\ref{results}, on results and discussion,  and Sec.\ref{exp} deals with certain possible experimental realizations and heating issues related to our model. We see that, the kicking scheme incorporates itself into the effective Hamiltonian in a way that it provides a means of manipulating the inversion symmetry breaking 
parameter of the Haldane model which is essentially the staggered off-set to the on-site energies at the two closest neighboring sites of the two interpenetrating triangular sub-lattices
$A$ and $B$. This and various other aspects are discussed therein, followed by a conclusion comparing our work to related studies. 

 The kicking protocol studied in this paper  has the effect of
 breaking inversion symmetry and opening a gap in graphene. This, however, does not lead to the appearance of any non-trivial topological features in graphene
 as the equal gap/mass term at the two Dirac points is of the Semenoff kind. We observe that in our present problem the general effect of the driving is to
 modify the inversion breaking energy of the Haldane model.
 
\section{The Kicked Haldane Model}
\label{Halmod}
\subsection{ The Undriven Haldane Model} 
The Haldane model \cite{HaldanePRL88} is a perfectly 2-dimensional Quantum Hall insulator with the unique property of
exhibiting Quantum Hall behavior in the absence of any net magnetic field through any of its unit cells. 
%It is one of the most elementary realizations of a Chern insulator as its band topology, under certain conditions, can be shown to belong to a non-trivial first Chern class. The way Haldane envisioned 
It consists of a  2-D hexagonal lattice of atoms,
 with a single tight binding orbital at each of the two lattice sites within a unit cell.  These are the two distinct sites   
\begin{figure}[b]
 \includegraphics[height=6cm,width=8cm]{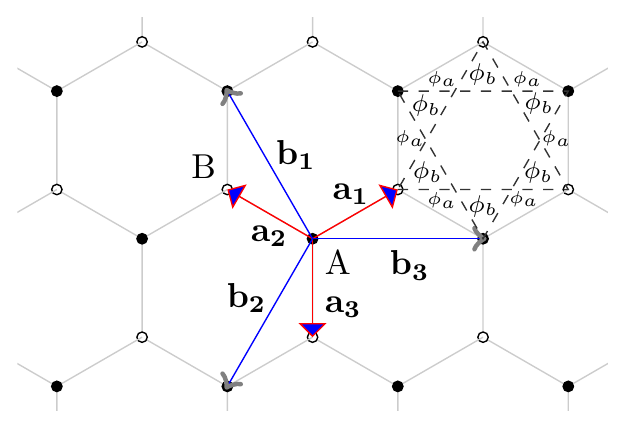}
 \caption{ Schematic illustrating the honeycomb lattice with vectors to the nearest and next nearest neighbors of an $A$ sub-lattice site. A flux configuration, where $\phi_a$ 
 and $\phi_b$ are fluxes through the triangles that bound each of the labeled regions, is
 also shown which can make the next to nearest neighbor hoppings phase dependent with zero net flux in the hexagon i.e. $(\phi_a+\phi_b)=0$. }
\label{fig:lattice}
 \end{figure}
belonging to the $A$ and $B$ triangular sub-lattices as shown in Fig. (\ref{fig:lattice}) by filled and hollow points respectively. Normally, such a lattice shows a semimettalic band structure which is 
well known from Graphene. However, to realize an insulator, the degeneracies at the Dirac points in the 2-D Brillouin zone need to be lifted by breaking the
inversion and time-reversal symmetries in the system. In the Haldane model these are broken to ensure Quantum Hall behavior. The inversion symmetry is broken by giving an off-set to the on-site energies at the two in-equivalent nearest neighbor sites $A$ and 
$B$ by an amount $-M$ and $+M$ respectively. Breaking inversion symmetry opens a gap at the band touchings in the Brillouin zone and makes the system a 
semiconductor/normal insulator. In order to get a topological insulator it is further required to break  time-reversal symmetry which is done here by making the 
hoppings to the next to nearest-neighbor sites complex valued as, $t_2e^{\pm i\phi}$, $t_2$ being real. 
The nearest neighbor hoppings $t_1$ on the other hand remain real valued. 
An ingenious choice of magnetic field helps to ensure this by making the overall magnetic flux through any of the hexagons (unit cells) of the lattice  zero and hence realizes a globally vanishing magnetic field
while at the same time breaking time-reversal symmetry. Though this does require the local existence  of a spatially periodic magnetic field everywhere, perpendicularly 
applied to the lattice plane. It gives  rise to a flux arrangement that collectively disappears over a unit cell. One such choice is illustrated in
Fig. (\ref{fig:lattice}), where the condition $(\phi_a+\phi_b)=0$ fulfills this requirement.  Several  such  choices are permitted by gauge freedom and 
since travelling along the sides of any hexagon encloses zero flux the $t_1$ hoppings acquire no phase contribution. The hopping term $t_2$ for the 
next to nearest neighbor sites acquires phases in hops around triangular cells which enclose non-zero flux. For the case in Fig. (\ref{fig:lattice}) this phase $\phi$ comes 
out to be $2\pi(2\phi_a+\phi_b)/\phi_0$ expressed in units of the flux quantum $\phi_0$.  

The need to break time-reversal invariance arises from the familiar requirement encountered in the IQHE \cite{QHE,Laughlingauge,TKNN,Halperinedge} that for non-zero quantized transverse conductance $\sigma_{xy}$, time-reversal invariance must
be absent in the system as otherwise $\sigma_{xy}$ is an odd function and amounts to zero. It is the behavior of the gap that opens at the Dirac points also called the mass term 
from the low energy, (2+1)-D relativistic linearization approximation, that crucially determines the existence of the Hall conductance. In the presence of just broken inversion symmetry
this mass term ($M$) is a Semenoff mass which has the same sign at both Dirac points and yields $\sigma_{xy}=0$ which follows from the definition for the 
Chern invariant, in this case. However if time reversal invariance is absent the mass term ($\phi$ dependent) has opposite signs at these points and leads to a non-zero $\sigma_{xy}$.When both parameters $M$ and $\phi$ are zero the bands touch at points
in the Brillouin zone called Dirac points owing to the linear dispersion in the vicinity of these degenracies. These are high symmetry points in the Brillouin zone
in addition to the band center. In this situation the system is semi-metallic and allows a two-dimensional representation at these high symmetry points. When 
the symmetry breaking parameters take on other combinations of values the system is found to belong to insulating regions with the Chern number $\mathcal{C}$, for the valence band (lower band
with the Fermi energy in the gap at zero temperature)
taking values $\pm 1,0$ depending on the relative strengths of the two parameters. These regions of different conductance values $\sigma_{xy}= \mathcal{C}e^2/h$ are 
separated by a boundary where the gap closes at either one of the Dirac points in the Brillouin zone. These touchings are the transition band configurations 
where $\mathcal{C}$ 's for the two bands can rearrange themselves by assuming values that add up to zero thereby ensuring the standard requirement that the total band bundle remains 
topologically trivial \cite{AvronSeilerPRL83}. These and other properties of the model follow from its Hamiltonian and the linear approximation to it at the band touchings.
We now take a closer look at this Hamiltonian which will serve as the target system for the intended driving scheme.

The two dimensional Haldane Hamiltonian in reciprocal space, as obtained from its real space tight binding form is
\begin{widetext}
\begin{equation}
\label{Halham}
H(\mathbf{k})= 2\mathbf{I}t_2\cos\phi\sum_i\cos(\mathbf{b_i}\cdot\mathbf{k})+ t_1\left[\sum_{i}\{\sigma^x\cos(\mathbf{a_i\cdot k}) 
 +\sigma^y\sin(\mathbf{a_i\cdot k})\}\right] + \sigma^z\left[M- 2t_2\sin(\phi)\sum_i\sin(\mathbf{b_i}\cdot\mathbf{k}) \right]
 \end{equation}
 \end{widetext}
Here, the quasi-momentum $\mathbf{k}$ is a good quantum number since the choice of magnetic field preserves the 
original translation symmetry of the lattice, $\mathbf{I}$ is the $2\times2$ identity element and $\sigma^x,\sigma^y$ and $\sigma^z$ are the Pauli matrices. From Fig.(\ref{fig:lattice}) 
the vectors $\mathbf{a_1}\equiv\left(\tfrac{\sqrt{3}a}{2},\tfrac{a}{2}\right)$, $\mathbf{a_2}\equiv\left(\tfrac{-\sqrt{3}a}{2},\tfrac{a}{2}\right)$ and $\mathbf{a_3}\equiv(0,-1)$
are the vectors from an $A$ sub-lattice site to the nearest neighboring $B$ sub-lattice sites. The $a$ stands for the length of the bond joining nearby $A$ and $B$ sites. This choice is a matter of convention here and is defined by these vectors forming a
closed right handed system with the cross product of any two in increasing sequence of the indices being aligned out of the plane in the direction of positive $\hat{z}$.  While, as seen in the same figure, the vectors to the next nearest neighbor sites are chosen 
as $\mathbf{b_1}=\mathbf{a_2}-\mathbf{a_3}$, $\mathbf{b_2}= \mathbf{a_3}-\mathbf{a_1}$ and $\mathbf{b_3}=\mathbf{a_1}-\mathbf{a_2}$. Thus the summation index in the 
above Hamiltonian extends over these three possibilities for both kinds of vectors. The reciprocal space lattice for this system is also  hexagonal and therefore 
first Brillouin zone (FBZ) is a hexagon with band touchings occurring at the zone corners. The FBZ comprises of two in-equivalent band touchings or Dirac points $\mathbf{K}$ and
$\mathbf{K'}$. It is possible to rearrange this hexagonal FBZ into an equivalent rhomboidal one by shifting regions of the former by reciprocal lattice vectors. Within
this description the Dirac points lie inside the FBZ and are given by $\mathbf{K}= \left(\tfrac{2\pi}{3\sqrt{3}a},\tfrac{2\pi}{3a}\right)$ and $\mathbf{K'}=\left(\tfrac{4\pi}{3\sqrt{3}a},0\right)$.
The two band energy dispersion that follows from the above Hamiltonian is 
\begin{widetext}
\begin{equation}
\label{Haleigval}
\begin{split}
  &E^{\rm H}_{\pm}(\mathbf{k}) = 2t_2\cos(\phi)\left[2\cos\left(\frac{3ak_y}{2}\right)\cos\left(\frac{\sqrt{3}ak_x}{2}\right)+\cos(\sqrt{3}ak_x)\right] 
 \pm \Biggl\{t_1^2\left[2\cos\left(\frac{ak_y}{2}\right)\cos\left(\frac{\sqrt{3}ak_x}{2}\right)+\cos(k_y)\right]^2 \\
 &+ t_1^2\left[2\sin\left(\frac{ak_y}{2}\right)\cos\left(\frac{\sqrt{3}ak_x}{2}\right)-\sin(k_y)\right]^2 + \Biggl[M -2t_2\sin(\phi)
 \times\biggl(-2\cos\left(\frac{3ak_y}{2}\right)\sin\left(\frac{\sqrt{3}ak_x}{2}\right)+\sin(\sqrt{3}ak_x)\biggr)\Biggr]^2\Biggr\}^{\frac{1}{2}}
\end{split}
\end{equation}
\end{widetext}

On substituting the coordinates for either of the Dirac points $\mathbf{K}$ or $\mathbf{K'}$ in the above expression one gets $M-3\sqrt{3}t_2\sin(\phi)$
and $M+3\sqrt{3}t_2\sin(\phi)$ respectively. From this we arrive at the condition for the bands to touch at these points as $M=3\sqrt{3}\nu t_2\sin(\phi)$ where
$\nu =\pm 1$ depending on the particular Dirac point under consideration. Touching at both points occurs only when both inversion and time-reversal are present
i.e both $M$ and $t_2\sin(\phi)$ are zero. The touchings at individual Dirac points occur in Haldane's Chern number phase diagram at the transition boundaries 
where $\mathcal{C}$ undergoes a discrete step in its value. 

An important aspect of the model is the constraint on the relative strengths of the hopping parameters given by
$\vert t_2/t_1\vert < 1/3$ that ensures that the bands of the model do not overlap. This is useful for a clear observation of the band touchings in any 
physical realization of the model as it ensures that the upper and lower bands are always well separated by a gap unless they touch with the energies at these
touchings being extremal points or maximas if one considers the lower band. We will discuss this condition in the context of  kicking later on to see how it
gets modified for the kicked system and also ascertain how it may be used to define a magnitude scale for the strength of the driving. Now, we move on to the model
of interest in the present work which is the Haldane Hamiltonian under kicking.

\subsection{ Driven Haldane Model}
\label{kickhalmod}
The choice of driving the Haldane model using a periodic train of delta function kicks allows an exact Floquet treatment of the stroboscopic kind
without recourse to a high frequency approximation of the kind used in \cite{InoueTanakaPRL10,WangLiWuEPL12}. Central to such approaches, and the marked rise of interest in Floquet topological
insulators, is the possibility of having a controllable parameter whose variation helps to tune the system from a normal to a topological insulator,
or through different topological phases. Thus a system may be designed where, by sweeping an experimentally controllable parameter across a prescribed range of values, one 
could transition the total Chern number of the filled bands of the system between trivial and non-trivial values. Much like the different quantized conductance
values assumed by the system in the IQHE when the magnetic field is swept adiabatically.  The added advantage driving has to offer here, is that it achieves 
all this in relatively simpler, non-interacting effective static Hamiltonians. Since, in general, topological characteristics are robust features of a system and are unaffected by perturbations to a large extent, having systems which do show transitions from normal to topological insulators (and vice-versa) in a
discrete manner is of considerable interest. This is so because interesting properties of the valence band Bloch functions are known to occur at the 
transitions such as, lack of a maximally localized Wannier representation in the Chern insulating phase and anomalous localization behavior of the wavefunctions
\cite{ThonhauserVanderbiltPRB06,SoluyanovVanderbiltPRB11}. Thus the transitions merit some attention in various systems where they can be realized in a manner
which permits a simpler analytical/numerical approach to their study. Our kicked model belongs to this category of systems.

Prior to expressing the Hamiltonian in the presence of kicking it would be useful to adopt some notation to denote terms in  Eqs. \eqref{Halham} and \eqref{Haleigval}.
The structure of the Hamiltonian in Eq.\eqref{Halham} is of the general form $H(\mathbf{k})= h_0(\mathbf{k})\mathbf{I}+ \mathbf{h}(\mathbf{k})\cdot\boldsymbol{\sigma}$, where
$\boldsymbol{\sigma} = (\sigma^x,\sigma^y,\sigma^z)$ is the vector of Pauli matrices and 
\[
h_0(\mathbf{k}) = 2t_2\cos(\phi)\left[2\cos\left(\tfrac{3ak_y}{2}\right)\cos\left(\tfrac{\sqrt{3}ak_x}{2}\right)+\cos(\sqrt{3}ak_x)\right]. 
\]
 The $\mathbf{h}(\mathbf{k})$ here, is the vector $[t_1L(\mathbf{k}),t_1F(\mathbf{k}),M-2t_2\sin(\phi)N(\mathbf{k})]$
with 
\[
L(\mathbf{k})=2\cos\left(\tfrac{ak_y}{2}\right)\cos\left(\tfrac{\sqrt{3}ak_x}{2}\right)+\cos(k_y)
\]
\[ 
F(\mathbf{k})=2\sin\left(\tfrac{ak_y}{2}\right)\cos\left(\frac{\sqrt{3}ak_x}{2}\right)-\sin(k_y) 
\]
\[
N(\mathbf{k})=-2\cos\left(\tfrac{3ak_y}{2}\right)\sin\left(\tfrac{\sqrt{3}ak_x}{2}\right)+\sin(\sqrt{3}ak_x)
\]. 

It follows that 
\[
\vert \mathbf{h}(\mathbf{k})\vert= \sqrt{t_1^2L^2(\mathbf{k})+t_1^2F^2(\mathbf{k})+(M-2t_2\sin(\phi)N(\mathbf{k}))^2}.
\]
The driving scheme is chosen to be a train of delta function kicks which are separated by fixed time interval $T$. Such a scheme was introduced in the context of driving a 
hexagonal lattice, in particular graphene, as a
platform for synthesizing novel dispersion relations and wave packet
dynamics \cite{AgarwalaDiptimanPRB16}. This work proposes using a kicking which is applied as the following perturbing term to the Hamiltonian
\begin{equation}
\label{Kicking}
             \mathcal{H}_{kick,\mathbf{k}}(t) =  (\alpha_x\sigma^x + \alpha_y\sigma^y + \alpha_z\sigma^z)\sum_{m=-\infty}^{m=\infty}\delta (t - mT)
\end{equation}
and represents a general $2\times2$ kicking protocol with the $SU(2)$ pseudo-spin structure of the 2-dimensional Haldane Hamiltonian. The $\alpha_x$, $\alpha_y$ and
$\alpha_z$ stand for kicking amplitudes in the respective directions. Since we are 
consistently expressing the Hamiltonian and the perturbation to it in $\mathbf{k}$-space, the kicking is applied uniformly to 
every unit cell of the lattice to have the reciprocal space representation of the above form. The dynamics 
of the system over a period $T$, under such a perturbation, are governed by an evolution operator $U_{XYZ} = U_{kick}U_{static} = e^{-i\boldsymbol{\alpha}.\boldsymbol{\sigma}}e^{-i\mathcal{H}(\mathbf{k})T}$
where, $U_{XYZ} = e^{-i\mathcal{H}_{XYZ}(\mathbf{k})T}$ with $\mathcal{H}_{XYZ}(\mathbf{k})$ as the Floquet Hamiltonian and, $\boldsymbol{\alpha}$ and $\boldsymbol{\sigma}$ are $(\alpha_x,\alpha_y,\alpha_z)$ and $(\sigma^x,\sigma^y,\sigma^z)$ respectively. Using the algebra of Pauli matrices and some standard 
results associated with them, it is possible (as illustrated in \cite{AgarwalaDiptimanPRB16}) to obtain the exact form of $\mathcal{H}_{XYZ}(\mathbf{k})$, see appendix.
 In particular, we are interested in a kicking scheme where $\alpha_z\neq0$ while $\alpha_x=\alpha_y=0$ and henceforth assume these parameter values
in the perturbing Hamiltonian in eq.\eqref{Kicking}. Thus we are interested in the $\hat{z}$-kicked Haldane model whose Hamiltonian we denote $\mathcal{H}_{Z}(\mathbf{k})$ 
which is obtained from $\mathcal{H}_{XYZ}(\mathbf{k})$ by putting in the requisite conditions. The calculation of $\mathcal{H}_{XYZ}(\mathbf{k})$ in the manner 
outlined in \cite{AgarwalaDiptimanPRB16} will involve considering only the vector $\mathbf{h}(\mathbf{k})$ projected along the Pauli matrices. The diagonal part
due to $h_0$ remains unmodified and finally shows up in the expression for $\mathcal{H}_{Z}(\mathbf{k})$ which is again of the structure $h_0(\mathbf{k})\mathbf{I}+ \epsilon_z(\mathbf{k})\mathbf{h'}(\mathbf{k})\cdot\boldsymbol{\sigma}$.
The vector $\mathbf{h'}(\mathbf{k})$ is represented by components $(h'_x(\mathbf{k}),h'_y(\mathbf{k}),h'_z(\mathbf{k}))$ which are
\begin{widetext}
\begin{align}
 h'_x(\mathbf{k}) = \frac{1}{\sin(T\epsilon_z)}\Biggl [\frac{-t_1L(\mathbf{k})}{\vert\mathbf{h}(\mathbf{k})\vert}\sin(T\vert\mathbf{h}(\mathbf{k})\vert)\cos(\alpha_z) 
 + {\rm sgn}(\alpha_z)\frac{t_1F(\mathbf{k})}{\vert\mathbf{h}(\mathbf{k})\vert}\sin(\alpha_z)\sin(T\vert\mathbf{h}(\mathbf{k})\vert)\Biggr] \nonumber
\end{align}
\begin{align}
 \label{KickedHalcomp}
 h'_y(\mathbf{k}) =\frac{1}{\sin(T\epsilon_z)}\Biggl [\frac{-t_1F(\mathbf{k})}{\vert\mathbf{h}(\mathbf{k})\vert}\sin(T\vert\mathbf{h}(\mathbf{k})\vert)\cos(\alpha_z) 
 - {\rm sgn}(\alpha_z)\frac{t_1L(\mathbf{k})}{\vert\mathbf{h}(\mathbf{k})\vert}\sin(\alpha_z)\sin(T\vert\mathbf{h}(\mathbf{k})\vert)\Biggr]
\end{align}
\begin{align}
 h'_z(\mathbf{k})= \frac{1}{\sin(T\epsilon_z)}\Biggl [-{\rm sgn}(\alpha_z)\sin(\alpha_z)\cos(T\vert\mathbf{h}(\mathbf{k})\vert)
  - \frac{M-2t_2\sin(\phi)N(\mathbf{k})}{\vert\mathbf{h}(\mathbf{k})\vert}\sin(T\vert\mathbf{h}(\mathbf{k})\vert)\cos(\alpha_z)\Biggr] \nonumber
\end{align}
 The energy eigenvalues of $\mathcal{H}_{Z}(\mathbf{k})$, i.e the $\hat{z}$-kicked Haldane model, without the offset
due to the $h_0(\mathbf{k})\mathbf{I}$ term of the undriven Haldane model, denoted by $\epsilon_z$, is given by 
\begin{equation}
\label{Floqspec}
\epsilon_z(\mathbf{k}) = \pm \frac{1}{T}\cos^{-1}\Biggl[\cos(\alpha_z)\cos(T\vert\mathbf{h}(\mathbf{k})\vert)
 -\frac{{\rm sgn}(\alpha_z)}{\vert\mathbf{h}(\mathbf{k})\vert}(M-2t_2\sin(\phi)N(\mathbf{k}))\sin(\alpha_z)\sin(T\vert\mathbf{h}(\mathbf{k})\vert)\Biggr]
 \end{equation}
and ${\rm sgn}(\alpha_z)$ in both the equations above is the sign of $\alpha_z$ function. 
\end{widetext}
This completes a description of the model Hamiltonian we are interested
in. We now give a brief overview of the mathematical formalism that shall be used to compute the topological invariant for this model.
\section{Computing the Chern Invariant and Hall conductance}
\label{cherninv}
The Chern invariant or Chern number for 2-D systems is the topological invariant that captures and quantifies the topological non-trivialities 
associated with the bands of a periodic system. The general definition involves treating the Bloch functions of the filled bands in any solid
as defining a principal fibre bundle over the FBZ which is a torus. The Chern invariant is then calculated for any given band as the integral of the Berry 
curvature, which may be obtained from the Berry connection defined on this bundle over the FBZ \cite{BerryPRSLA84,TKNN,KohmotoAnnPhys85}. This integral may be written in the 
following manner
\begin{equation}
\label{Chernintegral}
 \mathcal{C} = \int_{\rm BZ}\mathcal{F}_{k_x,k_y}(\mathbf{k})dk_x\wedge dk_y
\end{equation}
where $\mathcal{F}_{k_x,k_y}$ is an antisymmetric tensor denoting a curvature 2-form, the Berry
curvature or field. Haldane's work \cite{HaldanePRL88} suggests a simplified route to calculating the Chern number for the various topological phases
by an effective linearization of the spectrum at a Dirac point where the gap is like a mass term and coefficient of the $\sigma^z$ matrix in the linearized
Hamiltonian around this point. The total Chern number for the lower band is then given by the signs of the masses at the two in-equivalent Dirac points in the
FBZ as 
\begin{equation}
\label{Halchern}
 \mathcal{C}= \frac{1}{2}\displaystyle\sum_{\nu=\pm 1}\nu~{\rm sgn}(m_\nu),
\end{equation}
where $m_\nu$ is the mass term at the corresponding Dirac point indexed by $\nu$. Both the expressions are demonstrably equivalent and one can in principle
derive eq.\eqref{Halchern} from eq.\eqref{Chernintegral}. In our calculations we use both methods to develop the Chern number phase diagram in the presence of 
driving. The integration is performed numerically to validate the Hall conductivity quantization expected from the second definition.

We intend here to give a brief overview of the mathematical formalism adopted by us to compute the Berry curvature required in the 
above integral. This formalism is based on the concept of Bargmann invariants \cite{MukundaSimon1,MukundaSimon2,BargmannBerrylink}. 
 It essentially involves the use of $U(1)$ invariant pure state density matrices $\rho=|\psi\rangle\langle\psi|$ denote physical states or rays in a complex
projective ray space. Then, the Bargmann invariants, are products of these density matrices, $\rho_1\rho_2\cdots\rho_j$ with the $j$ states forming the vertices
of a $j$-sided polygon in ray space. In more explicit terms a Bargmann invariant of order $j$ for a set of as many normalized states $|\psi_j\rangle$
such that $\langle\psi_j|\psi_{j+1}\rangle\neq0$, is
\begin{equation}
\label{Bargmann}
 \mathcal{B}^j(\psi_1,\cdots,\psi_j)= \langle\psi_1|\psi_2\rangle\langle\psi_2|\psi_3\rangle\cdots\langle\psi_{j-1}|\psi_j\rangle\langle\psi_j|\psi_1\rangle
\end{equation}
 The phase of the Bargmann invariant in Eq.\eqref{Bargmann} is obtained as,
\begin{equation}
\label{Berrytensor}
 \mathcal{F}_{\alpha\beta}(\mathbf{x}) = \frac{1}{2i}{\rm Tr}(\rho(\mathbf{x}) \bigl[\partial_\alpha\rho(\mathbf{x}), \partial_\beta\rho(\mathbf{x})\bigr])
\end{equation}
i.e. the Berry curvature.  The $\mathbf{x} = (x_1,x_2,\cdots,x_{2N-2})$ denotes
coordinates of points in ray space under some suitable
parametrization, ray space being $(2N-2)$ dimensional for an $N$-level
quantum system. In the case of lattice systems and Bloch functions
these coordinates are $\mathbf{k}$-space coordinates
$(k_x,k_y,\cdots)$. The indices $\alpha$ and $\beta$ run over the ray
space dimensions. It is interesting to note that one can recover the
customary expression for the Berry curvature, over the Brillouin zone,
for $2\times2$ systems with translational invariance of the kind
$H(\mathbf{k})= \boldsymbol{\sigma}.\hat{n}(\mathbf{k})$, which is in
general given by \[ \mathbf{\Omega}(\mathbf{k})=
\frac{1}{2|\hat{n}(\mathbf{k})|^3}\hat{n}(\mathbf{k}).[\partial_{k_x}\hat{n}(\mathbf{k})\times\partial_{k_y}\hat{n}(\mathbf{k})] \]
upon making the substitution
$\rho(\mathbf{k})=\frac{1}{2}(1+\boldsymbol{\sigma}.\hat{n}(\mathbf{k}))$
in Eq.\eqref{Berrytensor}, $\mathbf{k}$ serving the role of
$\mathbf{x}$. This is drawn from a general analogy to the
spin-$\frac{1}{2}$ Bloch sphere construction for $2$-level systems
with Dirac structure. We use Eq.\eqref{Berrytensor} with the same
analogy for our $\hat{z}$-kicked Haldane Hamiltonian $\mathcal{H}_{Z}(\mathbf{k})$. 
 
  The use of density matrices to compute Berry's phase and hence topological invariants has been extended to finite temperature systems with the introduction
 of Uhlmann's phase \cite{DelgadoPRL14,RivasDelgadoPRL14}. In this case, going beyond pure state density matrices, the use of mixed-state density matrices allows one to
 go to finite temperature topological phases and characterize them. At low temperatures the invariant from Uhlmann's phase converges to the Chern number. This
 invariant has even been used to compute the topological phase diagram for the Haldane model at finite temperatures \cite{ArovasPRL14}. As a seperate study, that we 
 would like to consider in the future, it would be interesting to look at the kicked Haldane model at finite temperatures where topologically non-trivial phases might
 exist and observe the interplay of kicking and temperature.
 \section{Results and Discussion}
 \label{results}
 We shall now take up the discussion on (1) the  range of driving parameters and their effects on the band structure, (2)   effects of periodic kicking on the topological properties of the Haldane model and (3) the modification to Haldane's overlap criterion due to kicking.

 \subsection{Range of Driving Parameters and Effects on Band structure}
 
 \begin{figure*}[t]
\begin{tabular}{lll}
\includegraphics[height=5.10cm,width= 6.0cm]{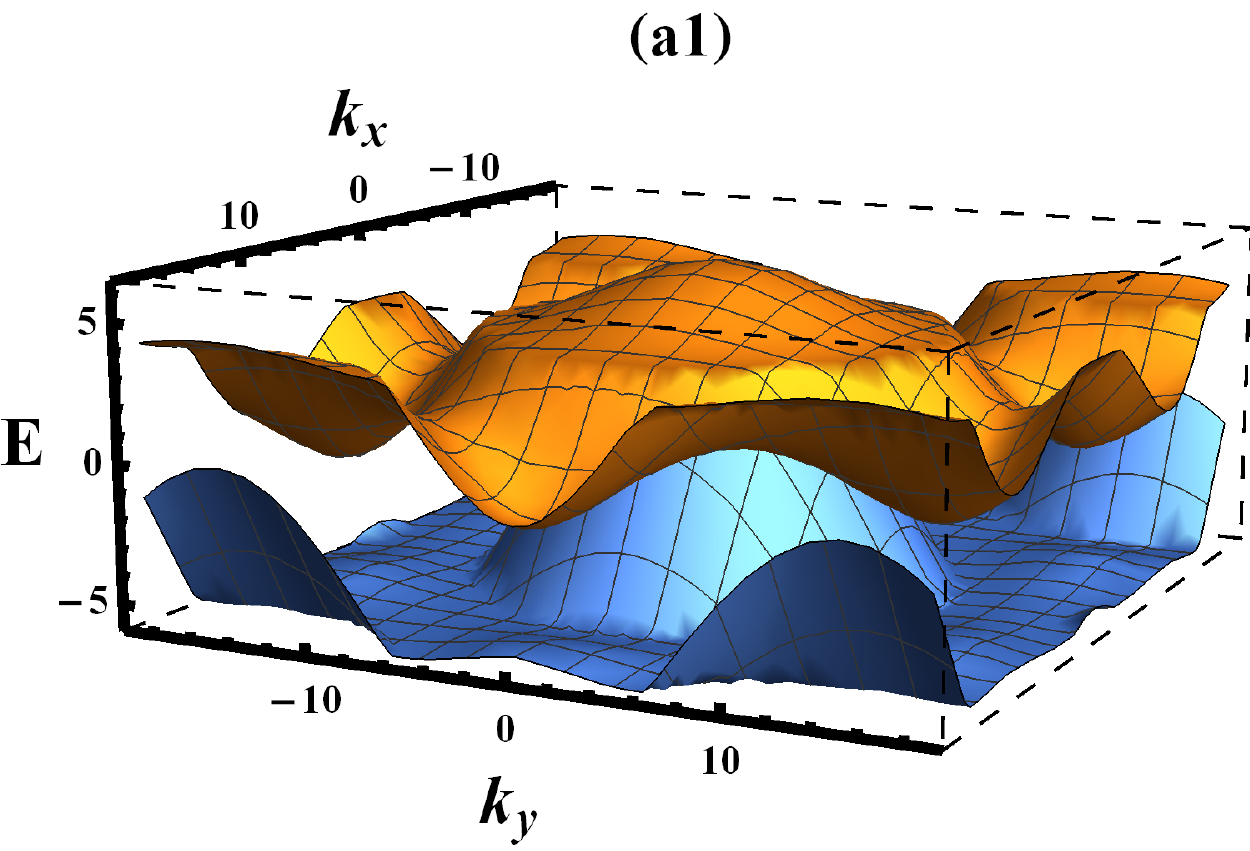}& \includegraphics[height=5.2cm,width= 6.0cm]{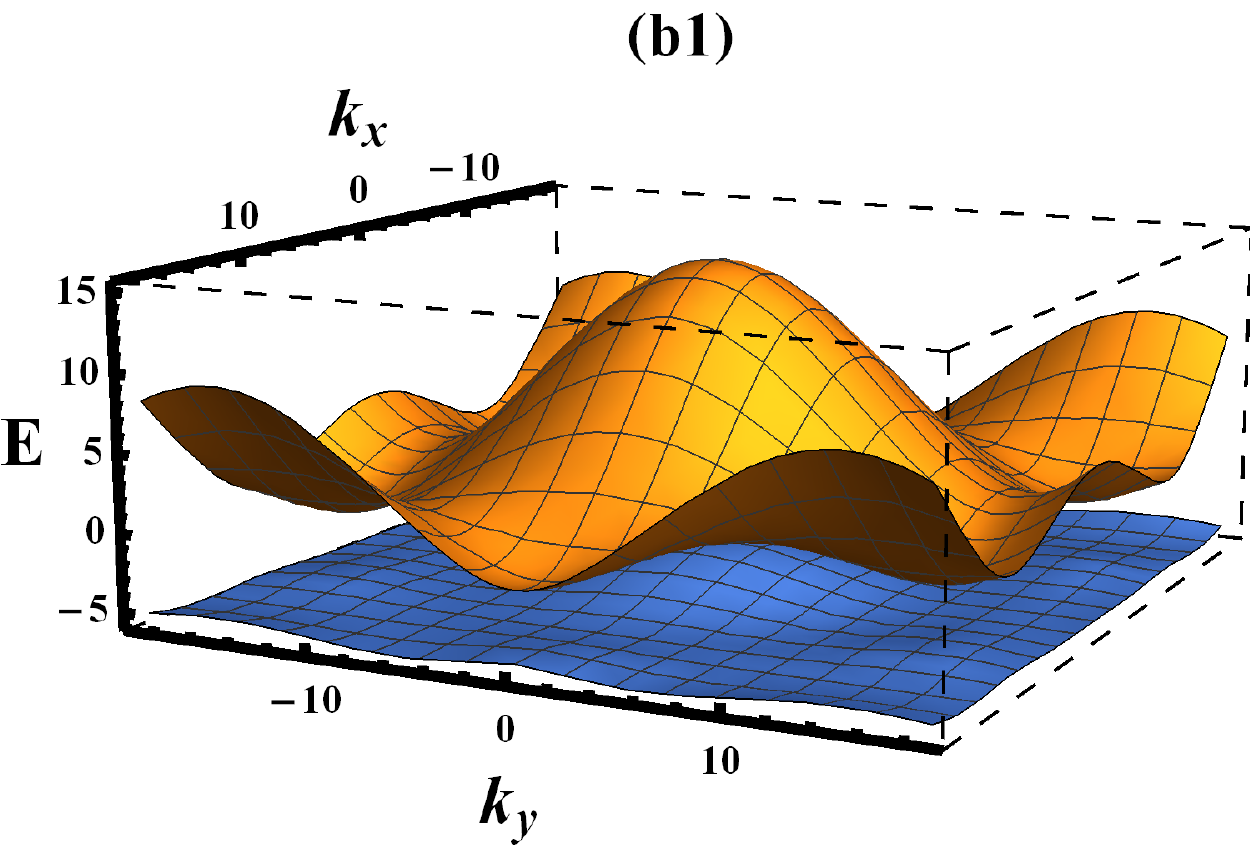}& \includegraphics[height=5.2cm,width= 6.0cm]{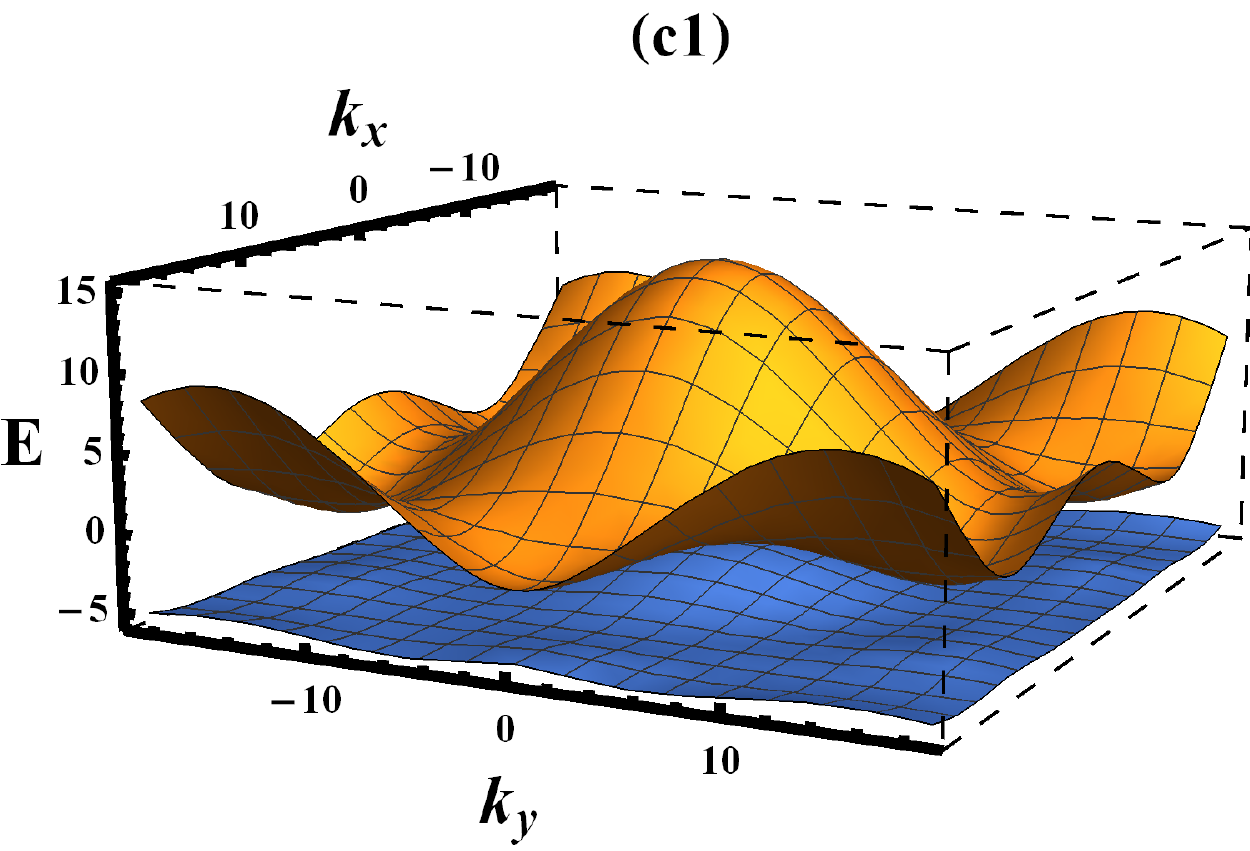} \\
\includegraphics[height=5.1cm,width=6.4cm]{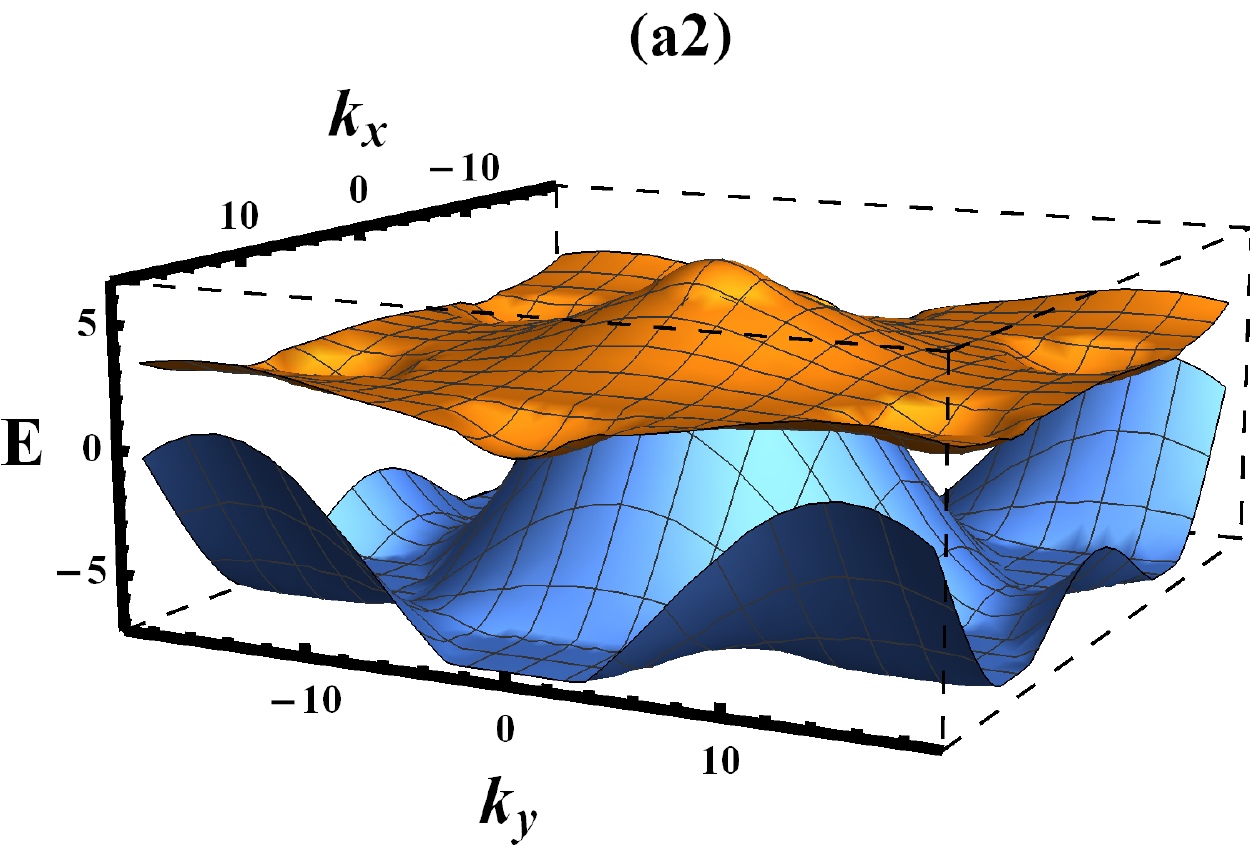}& \includegraphics[height=5.2cm,width=5.6cm]{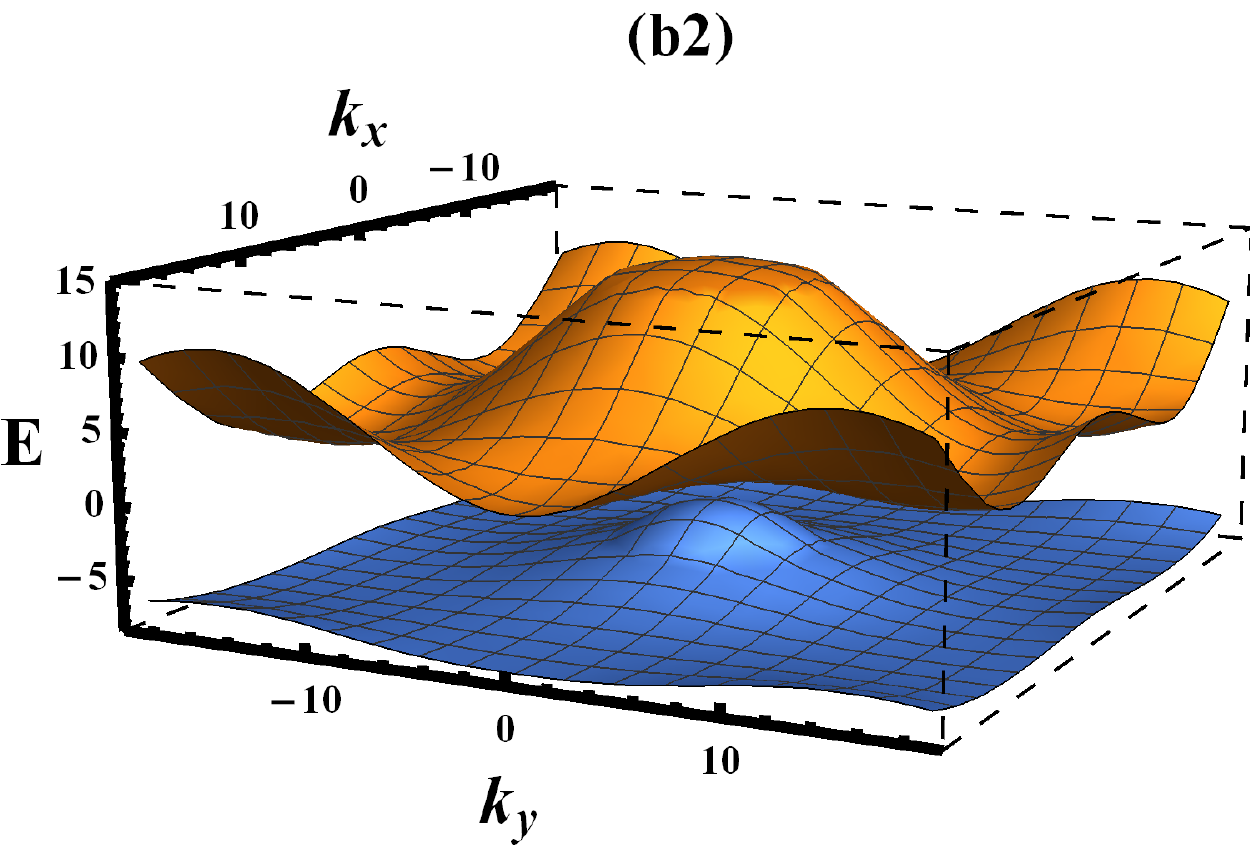}& \includegraphics[height=5.2cm,width=5.6cm]{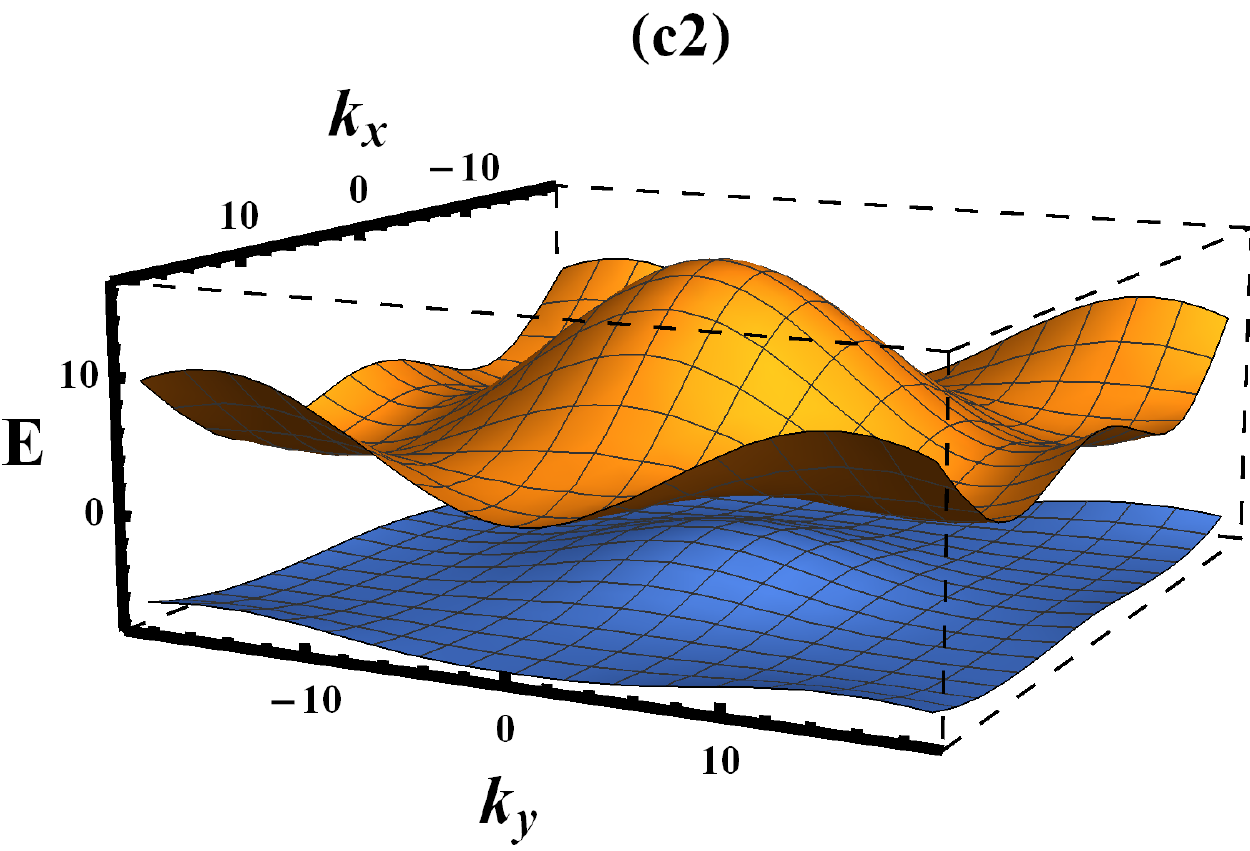}
\end{tabular}
\caption{  (Color online) The ground sate Floquet bands for the driven Haldane model as seen for different driving frequencies. We consider a driving period
of the form  $T=1/vt_1$ where $v=0.5,1$ and $2$. The fixed parameters for all plots are $\phi=0$, $t_1=3,t_2=1$ and $\alpha_z\approx0$. Plots (a1) and (a2) are for $v=0.5$ and
$M=2$ and $4$ respectively. In these we note that, for zero driving amplitude, the band structure is very much different from that of the undriven case. At
this low frequency the distortions are due to overlaps with Floquet sidebands. Plots (b1) and (b2) are for $v=1$ and $M$ values $2$ and $5$ respectively. Here, the 
choice of frequency works for the lower $M$ value without any higher band interference but for the larger $M$, sideband overlaps occur. This is seen from the
near flat truncation of the conductance band peak and the bump at the center of the valence band in plot (b2). Plots (c1) and (c2) are for $v=2$ and $M$ values $2$ and $5$ respectively.
They clearly indicate that this choice of frequency preserves the features of the Haldane spectrum when the driving is taken to zero.}
\label{fig:Fig2}
\end{figure*}

The Floquet Hamitlonian we have calculated is obtained 
 stroboscopically in an exact manner. Hence, there is in principle no restriction on the chosen driving frequency. However,  there are still bounds as to how low one can go.
  The behavior of the band structure of the driven model requires this lower limit to be set by  the convergence of the spectrum of the driven model to the undriven Haldane 
 spectrum in the limit of $\alpha_z\rightarrow0$ (i.e. taking the driving to zero).  
  We observe that one can go to a driving frequency of the order of energy $\approx t_1$, 
 if  the undriven Haldane model, has parameter values $t_2=1$ and $t_1=3$. This  choice
of parameters satisfies the overlap prevention requirement.
 
 To put this lower limit in perspective we note when $M =0$ and $t_2\sin(\phi) = 0$,  the bandwidth of the Haldane model is $\approx 6t_1$ and hence one can work with a frequency up to this order.  In this situation  neither inversion nor time-reversal are violated and the system allows bands to touch at both Dirac points in the FBZ. The presence of $M$ alters the bandwidth but is  of no substantial influence if considered smaller than the nearest neighbor hoppings $t_1$.  
 
 For larger $M>3.5$, there are overlaps
 of the ground state Floquet bands with the Floquet sidebands for driving period $\approx 1/t_1$. In this case it is observed that an upper limit to the driving period $T=1/2t_1$
 resolves this issue for all $M$ choices. The issue with larger $M$ s in the $1/t_1$ limit case can be resolved at non-zero driving amplitudes which remove the 
 overlap to the sidebands but this does not hold true when one goes all the  way down to zero driving amplitude. Thereby by making $1/2t_1$ the more favourable choice of 
 upper limit for the period. These features are illustrated in Fig.\ref{fig:Fig2}.

  So, in a driving scheme based on periodic kicking we are able to free the analysis of the  constraint of limiting the driving to high frequencies and instead go to comparatively lower values. This feature is absent in the schemes involving continuous drives, such as 
 circularly polarized light, that require the photons of the driving radiation to be of energies larger than the bandwidth  \cite{InoueTanakaPRL10}.
 
 This brings us to the question of how the amplitude of driving influences the features of the driven system. 
 %Presently, like in the discussion on driving  frequencies, 
 We restrict ourselves  to a discussion of how the driving amplitude affects the band structure for a fixed choice of the hopping energies and at some particular choice of $M$
 and $\phi$. The
 driving accentuates the inversion symmetry-breaking and the gap that opens in the spectrum increases as the amplitude is increased. There are however effects on the 
 band curvature. It is known that when the kicking  is applied to graphene, it leads to flat band structures at driving amplitudes of the magnitude $\alpha_z= \pi/2$ \cite{AgarwalaDiptimanPRB16}. 
 In the Haldane model one of the crucial differences in the band structure from that of ordinary graphene is the absence of particle hole symmetry (due to the 
 next nearest neighbor hoppings governed by $t_2$). This feature is loosely understood in terms of the greater number of $B$ sites than $A$ sites in any finite bounded
 version of the system.

 \begin{figure*}[t]
\begin{tabular}{ll}
\includegraphics[height=5.10cm,width= 6.0cm]{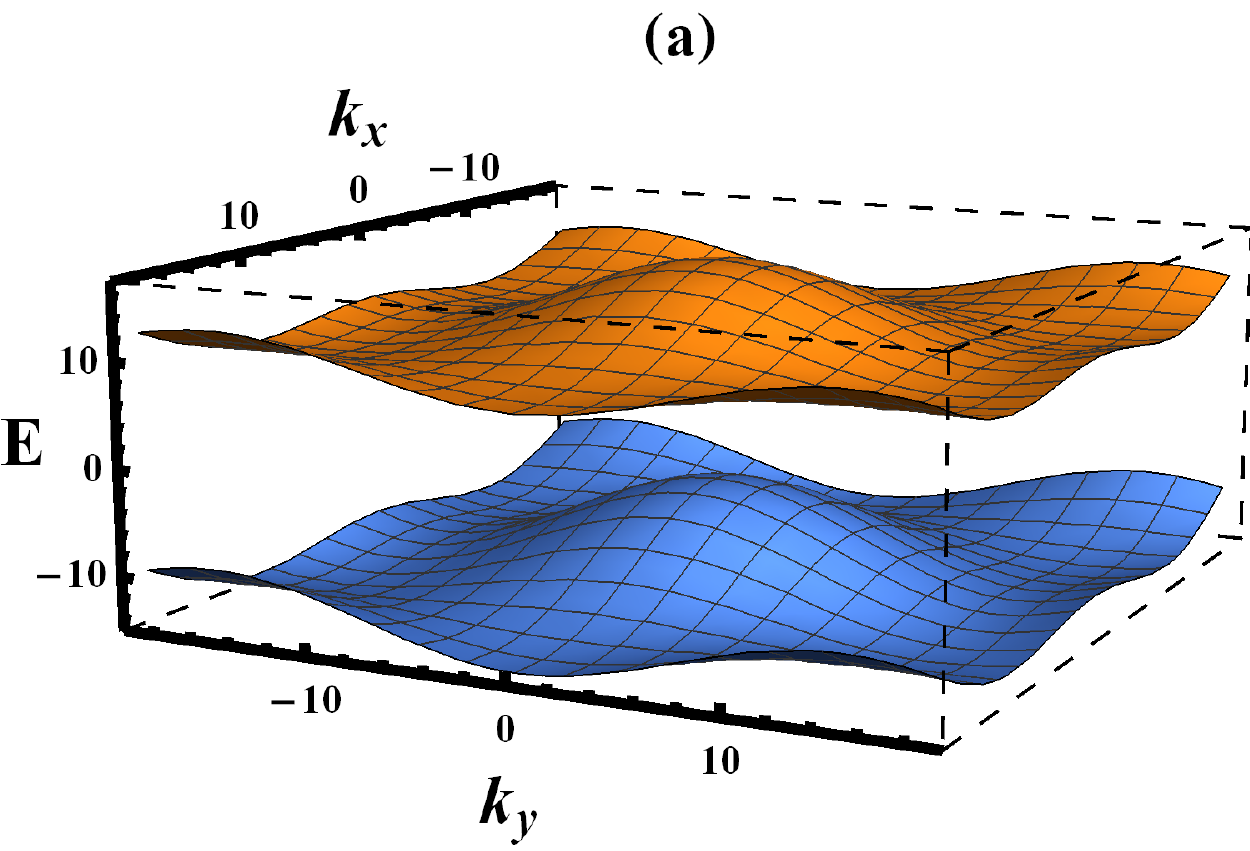}&
\includegraphics[height=5.2cm,width= 6.0cm]{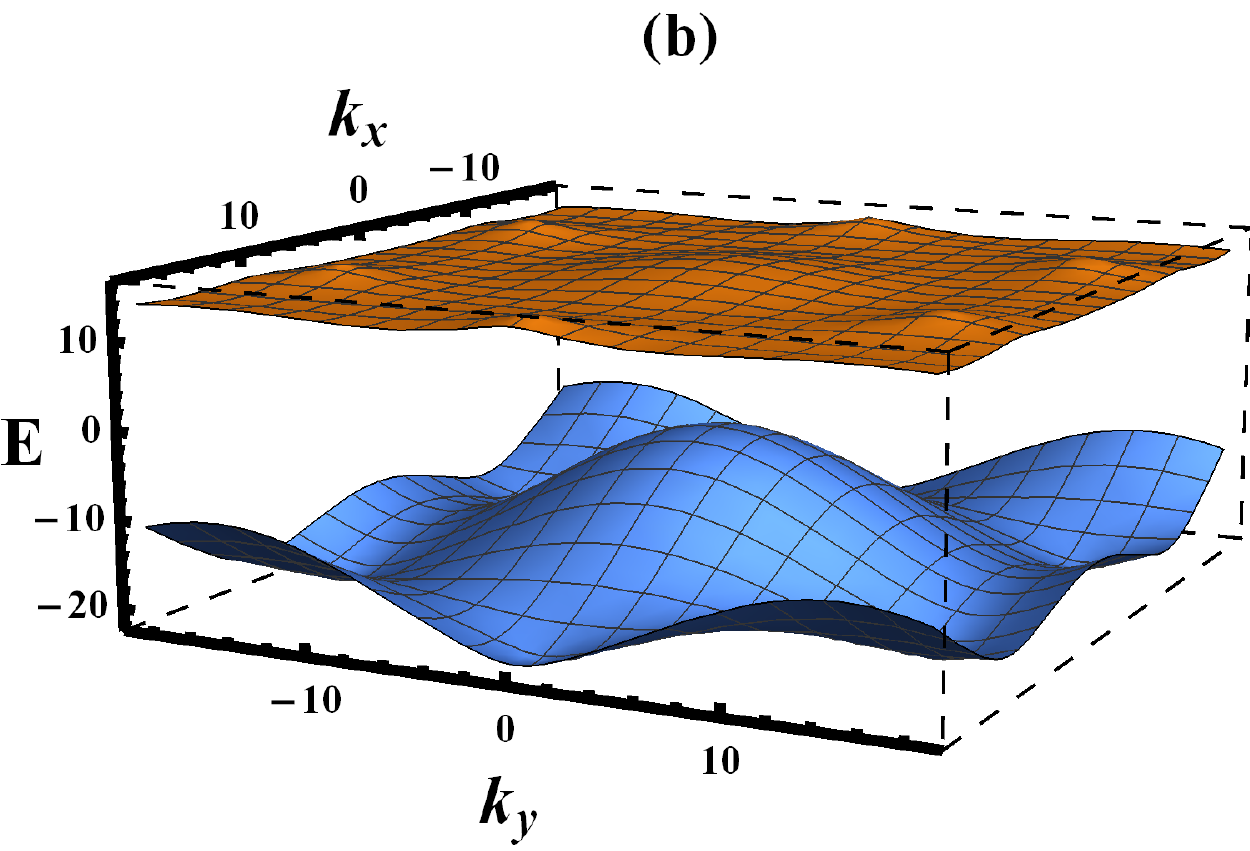}
\end{tabular}
\caption{ (Color online) The behavior of the Floquet spectrum at relatively large driving amplitudes for the fixed parameters $\phi=0$, $t_1=3$ and
$t_2=1$ . Plot (a) is for $\alpha_z=\pi/2$ and $M=2$. It shows the appearance of the tendency to flatness in the band structure especially in the conduction band. Changing
the $M$ value does affect this behavior but the flatness can be found to occur at a suitable corresponding $\alpha_z$ value. 
Plot (b) is for $\alpha_z=\pi$ and $M=0$. We see that the conduction and valence bands have completely exchanged their structures from the undriven case. Important to note
that the conduction band now shows almost complete flatness with slight bumps at the Dirac point locations which touch with the minima of the valence band when one
accounts for the folding of the Floquet quasienergies. Toplogically this has the effect of exactly exchanging the Chern numbers for the bands from the undriven case.
This is discussed in detail in Sec.\ref{topofeatures}.}
\label{fig:Fig3}
\end{figure*}
 
 Thus, for this model, when the amplitude of kicking is similarly increased, the band structure 
 does not  become  completely flat,  especially for the valence band. The conduction band does show nearly perfect flatness when the hopping 
 energies are in the ratio satisfying $\vert t_2/t_1\vert<1/3$.

 The choice of $\phi$ here is kept fixed at $0$ and $M$ could be non-zero but within the
 range that shows topological behavior in the undriven case, i.e $[-3\sqrt{3}, 3\sqrt{3}]$. 
  Though one may be cautioned that even in going upto this magnitude of driving the 
 undriven overlap condition begins to break down in favor of a newer one hinted at earlier, but the signatures of flatness can be observed to occur well 
 before this threshold is reached. In going beyond the $\alpha_z=\pm \pi/2$ limit the band structure is found to invert its curvature 
 and as one proceeds to increase the driving to $\alpha_z=\pi$ the conduction and valence exchange their structure from what is seen near zero driving. The interplay of the magnitude of $M$ 
 and $\alpha_z$ is found to effect the degree of flatness of the bands, especially the conduction band. These features are illustrated in Fig.\ref{fig:Fig3}. 
 We will see that due to the periodicity in the mass term stemming from the nature of the kicking, changing the magnitude of the driving causes  the system to undergo transition in and out
 of topological phases in a periodic manner.  In order to observe the full array of non-trivial topological behavior it suffices to work in the 
 driving amplitude range of $\alpha_z\in[-(2n+1)\pi/2,(2n+1)\pi/2]$ and further within this range, the original condition to avoid overlap of bands when touchings occur i.e. 
 $\vert t_2/t_1\vert<1/3$ is valid almost within $\alpha_z\in[-1,1]$. This range is sufficient to observe the competition between $M$ and the driving in terms of 
 influencing the topological phase, for a fixed choice of hoppings satisfying the above criterion.
 
 However,  to maintain sufficient generality in our discussion we will look at topological behavior at large driving amplitudes and the new overlap condition that comes 
 into play in these regimes. A point to note here is that though we fix hopping values while discussing the topological properties at large drivings (thereby 
 falling out of the criterion to avoid overlap of bands at these large driving amplitudes), this effect may be ignored so far as understanding  of the topological
 phases is concerned. If one is indeed interested in a realization of the driven model at high amplitude kicking and in observing the band touchings in the spectra,  an adjustment in the choice of hoppings, especially $t_2$, is necessary. Speaking in these terms necessarily assumes that one is working with a system where  such parameters as the hopping energies and the site energies are free to be controlled and varied. This seems possible only in optical lattice setups where
 lattice depths and occupation densities of the ultracold atoms can be manipulated.  
  
\subsection{Topological features of the kicked model}
 \label{topofeatures}
 \subsubsection{Analytical Deductions}
 We now come to a discussion of the topological properties of the driven Haldane model. Here, we analyze the effects of periodic kicking on the topological phase diagram  for the Hamiltonian in eq.\eqref{Halham} \cite{HaldanePRL88}. We look at the mass term of our driven model,  which is the coefficient of the $\sigma^z$ matrix in 2D systems, for the various topological phases the 
 system could exhibit. Thus we make use of the definition for obtaining the Chern number $\mathcal{C}$ given in eq.\eqref{Halchern}. To apply this we consider
 $\epsilon_z(\mathbf{k})h'_z(\mathbf{k})$ from eq.\eqref{KickedHalcomp}, which is the coefficient of $\sigma^z$ in the driven Haldane Hamiltonian. The technique requires one to consider the gap at the 
Dirac points $\mathbf{K}$ and $\mathbf{K'}$, and  look at the sign of $h'_z(\mathbf{k})$ in the vicinity of these points. On doing so $\mathcal{C}$ is given by
the expression 
\begin{widetext}
 \begin{equation}
 \label{Chernno}
 \begin{split}
 \mathcal{C}= \frac{1}{2}\displaystyle\sum_{\nu=\pm 1}\nu{\rm sgn}\Biggl[\frac{\epsilon_z(\mathbf{k})}{\sin(\gamma_\nu)}\biggl(-{\rm sgn}(\alpha_z)\sin\alpha_z
  \cos(T\vert M-&3\sqrt{3}\nu t_2\sin \phi\vert)\\
  &- {\rm sgn}(M-3\sqrt{3}\nu t_2\sin \phi)
 \sin(T\vert M-3\sqrt{3}\nu t_2\sin \phi\vert)\cos\alpha_z   \biggr)\Biggr]
 \end{split}
 \end{equation}
 
where, 
\begin{equation}
\label{denom}
\gamma_\nu= \cos^{-1}\biggl[\cos \alpha_z\cos(T\vert M-3\sqrt{3}\nu t_2\sin \phi \vert)- {\rm sgn}(\alpha_z)
   {\rm sgn}(M-3\sqrt{3}\nu t_2\sin \phi)\sin \alpha_z\sin(T\vert M-3\sqrt{3}\nu t_2\sin \phi \vert) \biggl]
\end{equation}
\end{widetext}
The denominator in the above expression for $\mathcal{C}$ goes to zero
for certain values of the driving $(\alpha_z,T)$ and the Haldane model parameters $(M,\phi,t_1,t_2)$. Out of these the hopping parameters will usually be
considered to be fixed for a given realization of the model. Here,  we are interested in the general conditions that can be deduced 
from the form of the Chern number and the behavior of the mass term at the Dirac points under various choices of the driving and  model
parameters. 

Thus the condition for the denominator to go to zero, the mass terms to vanish, and hence Berry curvature  to diverge at either of the Dirac points, is given by  $\gamma_\nu = n\pi$, with $n= 0,\pm 1,2,3\dotsc$. 
This essentially reduces to the condition $\cos(\vert\alpha_z\vert +  T\left ( M-3\sqrt{3}\nu t_2\sin \phi\right ))=\pm 1$ which implies 
$\vert\alpha_z\vert + T\left ( M-3\sqrt{3}\nu t_2\sin \phi\right )=n\pi$.
The numerator of the expression for $\mathcal{C}$  (see  eq.\eqref{Chernno}), apart from the $\epsilon_z(\mathbf{k})$ term which does not play a role in determining the sign of the term (at the locations for the
two Dirac points once one has chosen the valence band  for calculating $\mathcal{C}$), go to zero for $\sin(\vert\alpha_z\vert + T\left (  M-3\sqrt{3}\nu t_2\sin \phi \right )=0$ .  The appearance of the  indeterminate $0/0$ form which seems to occur is regulated 
in a limiting manner, by the presence of the $\epsilon_z(\mathbf{k})$ in the numerator. Thus what we have obtained is the condition for the bands to touch at either one of the Dirac 
points depending on the value of $\nu$ ($\pm 1$) in the equation $\vert\alpha_z\vert +  T\left ( M-3\sqrt{3}\nu t_2\sin \phi \right ) =n\pi$.

This is the modified condition for the boundary sinusoids which enclose the topologically non-trivial phases in the case of the  Haldane model under kicking. 
\begin{figure*}[t]
\begin{tabular}{ll}
\includegraphics[height=5.70cm,width= 6.2cm]{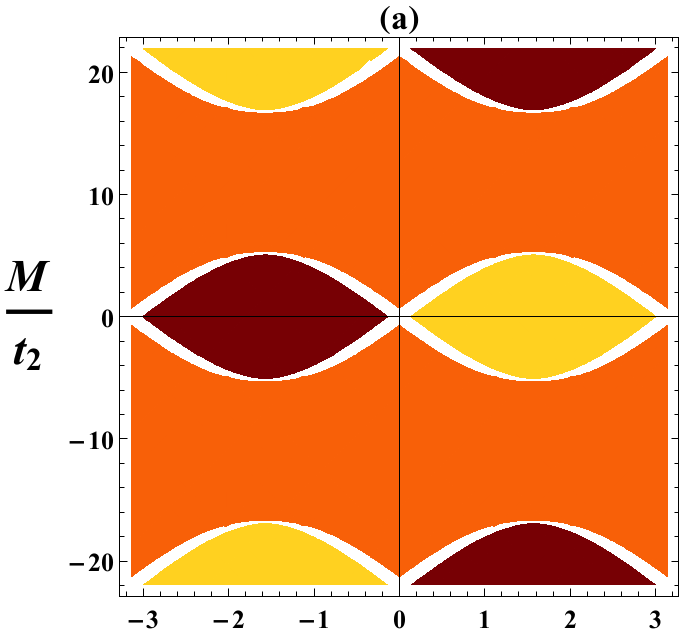}& \includegraphics[height=5.7cm,width= 6.3cm]{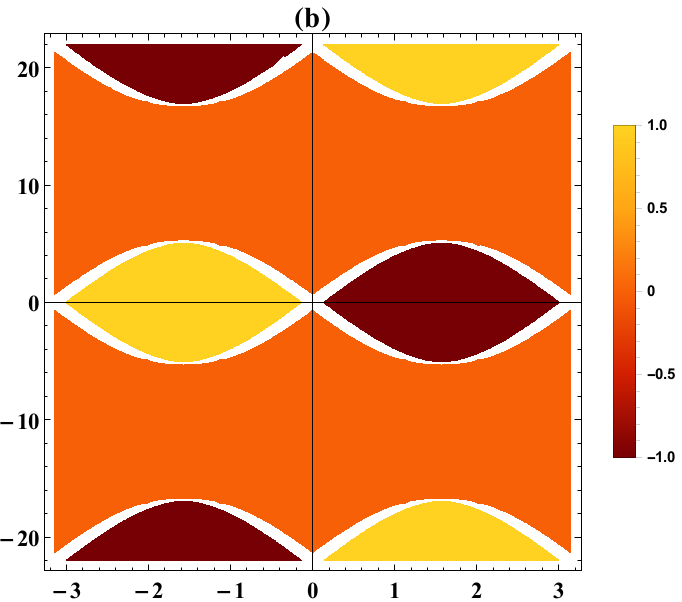} \\
\includegraphics[height=5.7cm,width=6.2cm]{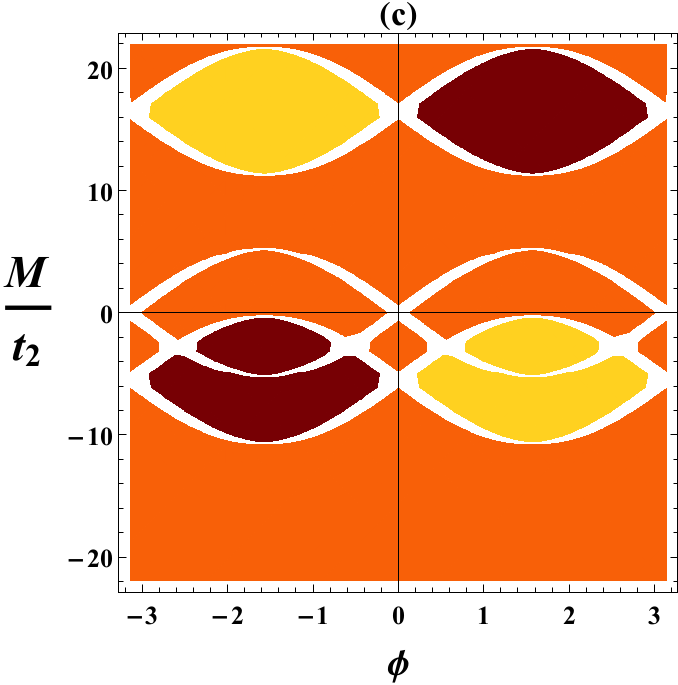}& \includegraphics[height=5.6cm,width=5.6cm]{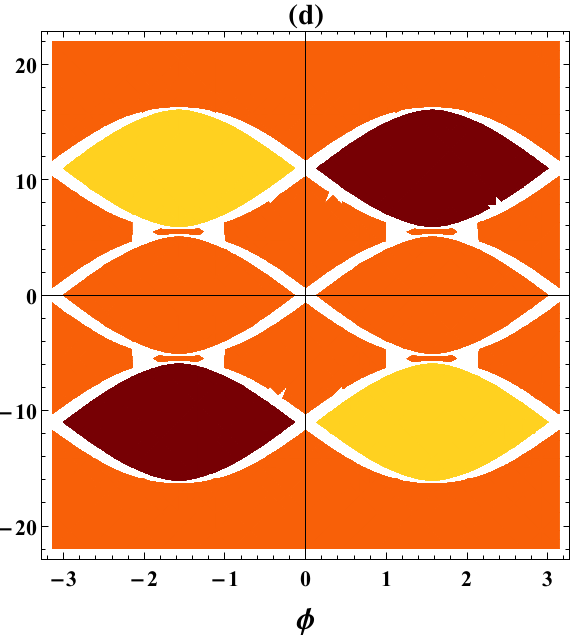}
\end{tabular}
\caption{ (Color online)Plots of the Chern number in the $M/t_2$ and $\phi$ plane for different $\alpha_z$ values. The plots (a),(b),(c) and (d) are for
$\alpha_z$ values $0~\textrm{{\bf (in the limit of weak driving)}}~,\pi,\pi/4$ and $\pi/2$ respectively. The darkest regions indicate a Chern number of $-1$, the lightest ones $1$ and the intermediate 
shade is for $0$ Chern numbers. The choice of undriven Haldane model hoppings, for these plots, is fixed at $t_1=3.5$ and $t_2=1$. The driving period is taken 
to be $T=\tfrac{1}{2t_1}$.}
\label{fig:fig2}
\end{figure*}
 
A couple of features become apparent from this condition. We observe, that the periodic kicking has the effect of modifying the inversion breaking parameter $M$ to $M-\tfrac{(n\pi-\vert\alpha_z\vert)}{T}$ which depends on the driving parameters $\alpha_z $ and $T$. Thus for different values of $n$, there is a specific set of values for $(M,\alpha_z,T)$ which would satisfy phase boundary conditions similar to the conditions satisfied by the  Chern number 
 in the Haldane model.  In this case we have a periodic recurrence of the phase diagram plotted between $M/t_2$ and $\phi$ along the $M/t_2$ axis, as manifested in repeated copies of the original Chern diagram for 
 the undriven model on moving along this axis. Thus, the broad topological behavior of the undriven model is preserved  in the driven model but now extends to
 newer regions of $M$ values for a fixed choice of $t_2$. The system under driving begins to explore a larger space of parameters 
 in terms of the occurrence of topological phases. Another feature that comes across is that the new condition for the phase boundaries depends on the magnitude of the  driving $\vert\alpha_z\vert$ and is independent of its sign. In fact, the modification to the inversion breaking factor is such that it depends on the ratio $\alpha_z/T$  which encapsulates the complete effect of the driving. The appearance of the ratio indicates that the amplitude of the driving can be made to scale
 with the frequency in a linear fashion to obtain a class of driven models with identical topological behavior. There is even the possibility of
 choosing the amplitude of the kicking to gradually increase, in a linear fashion with time, on a scale adiabatic in comparison to the driving, so as to be
 effectively regarded as constant over several driving periods. With this one may realize a linear-in-time variation of the inversion breaking term and hence 
 travel from a topologically non-trivial to a topologically trivial phase. This could be of use in schemes looking to quench Chern insulators across a 
 topological phase boundary with a normal insulator to study various properties of dynamical topological phase transitions at the quantum critical point\cite{DuttaUtsoPRB17}. 
  \begin{figure*}[t]
\begin{tabular}{lcr}
\includegraphics[height=5.70cm,width= 5.4cm]{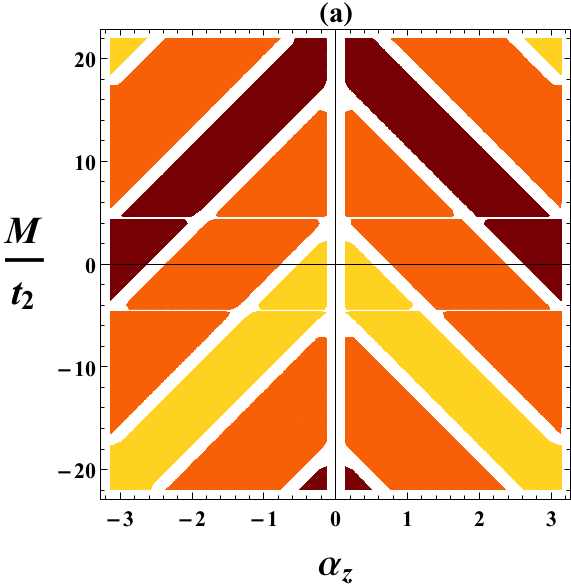}&
\includegraphics[height=5.7cm,width= 5.3cm]{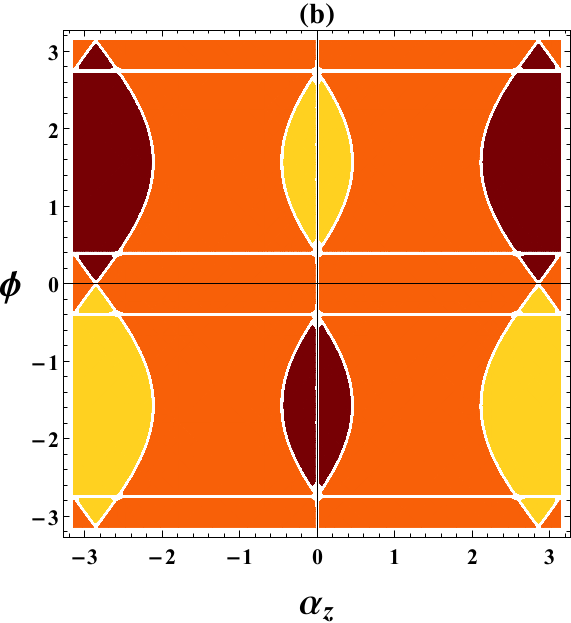}&
\includegraphics[height=5.7cm,width=5.6cm]{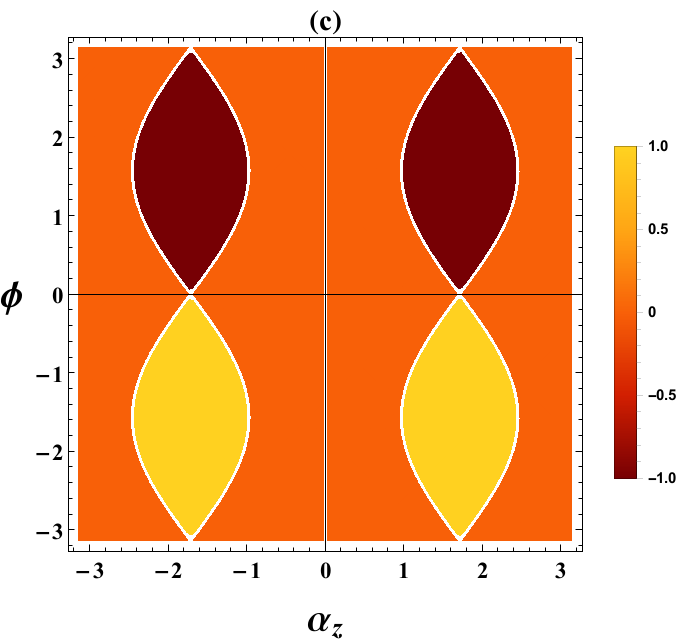}
\end{tabular}
\caption{ (Color online) Various sectional views that indicate how the topological phase diagram varies with change in driving amplitude keeping one symmetry
breaking parameter fixed and varying the other. Plot (a) indicates such variation for a fixed $\phi=\pi/3$ and illustrates the linear variation of the 
phases for different $M/t_2$ as $\alpha_z$ is changed. One notes the sharp turn the phases take at $\alpha_z\boldsymbol{\rightarrow}0$, a clear consequence of the dependence of 
the phases on the driving magnitude $\vert\alpha_z\vert$. Plots (b) and (c) illustrate phase regions for varying $\phi$ and $\alpha_z$, with $M/t_2$ choices
fixed at $2$ and $10$ respectively. The Chern number convention for the shaded regions is the same as that for Fig.\ref{fig:fig2}.}
\label{fig:fig3}
\end{figure*}
 
 The effect of increasing the driving amplitude from zero (in either positive or negative sense), i.e. the undriven situation, is to shift the Haldane Chern number phases (pair of lobes 
 due to the intersecting sinusoidal phase boundaries) vertically downwards, from their undriven position, along the $M/t_2$ axis. This effect applies to all the periodic copies of the phase diagram
 along this axis. Let $M'$ be used to denote the new effective inversion breaking parameter in the presence of driving.  Thus what we are effectively witnessing is a renormalization of the `Semenoff mass' component $M$ in the Haldane mass. 
 In the undriven case there was a unique inversion breaking site energy $M$  with the phase diagram center at $(M=0,\phi=0)$. This then had corresponded to a graphene like semi-metallic  band structure with touchings at both Dirac points.  In the driven model this admits multiple values as seen from $M'\equiv M-\tfrac{(n\pi-\vert\alpha_z\vert)}{T}$ and hence multiple semi-metallic centres $M-\tfrac{(n\pi-\vert\alpha_z\vert)}{T}=0,\phi=0$.
for the different $n$ and $\alpha_z$ values. 
The $n$ values define a set of several `Semenoff Masses' at a give non-zero kicking
all of which are valid choices around which topological phases can manifest. There is now a multiplicity of possible undriven Semenoff mass choices $M$ which yield  $M'=0$. The period of driving $T$, which we fix with a specific $t_1$, decides the separation
between the centres for a given driving. Thus the zero driving case does not collapse to a single $M$ value topological phase structure ( the original Haldane model)  but still shows a multitude of such phase diagrams which may be regarded as a consequence of the folding or periodicity
in Floquet quasienergies. This hints that the topological phase diagrams repeat identically at a separation of
$2\pi/T$ in the $M$ values which is exactly the width of a quasienergy Brillouin zone.

Varying the driving $\alpha_z$ on the other hand, for a fixed choice of $n$ and $M$, is a more physically plausible and interesting  as that would take a chosen undriven model $(M,\phi)$ through a topological transition.This is very much like quantum Hall plateau transitions with adibatically varying
magnetic field. An interesting feature that shows up is that for a given kicking amplitude at $M$ values $\tfrac{(n\pi-\vert\alpha_z\vert)}{T}$, for different $n$ say, $0$ and $1$
the Chern number phases are reflected about the $\phi=0$ line in the phase diagram. This is of more significance when one varies the driving amplitude $\alpha_z$ 
to the relatively high regime of $\pi$ or $-\pi$. Then the Semenoff mass $M'$ post driving is equal to the undriven one $M$ for $n=1$ which is clear from the relation. So the
Chern number phase diagram with its phases reflected about $\phi=0$ now occupies the region of the phase diagram where earlier the undriven Haldane Chern number
diagram was valid and thus in this extreme driving condition the topological phases undergo an exact inversion. This indicated  that even an inversion breaking taken to
a certain extreme may alter the band topology of a Chern insulator atleast in the presence of driving. However, physically there are issues with such 
large kicking amplitudes some of which have been discussed earlier. 

Again one has to exercise some caution here, on account of the folding of the quasienergies. 
There is always the possibility of band touchings which occur at the extreme ends of the spectrum (quasienergy Brillouin zone boundaries), besides the conventional ones at the middle of the spectrum 
which occur in the undriven and driven cases. This  could cause the Chern numbers to invert for the two bands. Indeed, what we see here is that, the inversion in phases is due to these band touchings at the
$\pm \pi/T$ limits of the folded spectrum and hence the gap closing at the edges of the quasienergy Brillouin zone. These arguments sit well with the 
previous discussion of the appearance of flat band behavior in the conduction band as driving amplitude is increased. As it starts to acquire the curvature characteristics which are present in the valence band in case of
zero driving. Thereby an exact reversal of structure occurs between the valence and conduction bands and the Chern numbers flip due to this new closing opening transition. The band structure is shown in  Fig. \ref{fig:Fig3}(b).

\subsubsection{Evidence from Phase Diagrams}
 To illustrate the various aspects of the topological phase diagram for the
 driven Haldane model we refer Fig.(\ref{fig:fig2}). These figures are for
 parameter values $t_1=3.5$ and $t_2=1$ that satisfy the band overlap
 prevention condition. Again, we caution that variation to this condition in
 the presence of driving which has been hinted on several earlier occasions
 and so $t_2$ has to be changed beyond a certain driving amplitude regime but
 here this is ignored as the broad topological behavior is unaffected by
 this. The driving period is fixed at $T=\tfrac{1}{2t_1}$. This choice, as
 stated earlier, ensures that the limits of the Floquet quasienergy Brillouin
 zone remain beyond the bandwidth of the undriven model and thus manifests in
 the phase diagrams as avoided overlaps between the different replicas of the
 intersecting sinusoids that are seen one below the other in
 Fig.(\ref{fig:fig2}). Other previously discussed features that become
 apparent include, for instance, one can look at the the plots in
 Fig.(\ref{fig:fig2}) (a) and (b) which are for $\alpha_z=0$ and
 $\alpha_z=\pi$ respectively and note that when the driving is taken to such
 extremes the band topology inversion spoken of earlier occurs. Additionally,
 though $\alpha_z$ is taken to tend to zero for the plot (a) and one
 does indeed see the undriven Haldane model phase diagram around the
 $(M=0,\phi=0)$ centre, there are still copies of similar non-trivial
 topological phases along the $M/t_2$ axis which are absent in the original
 Haldane model. This indicates that the stroboscopic Floquet Hamiltonian does
 not converge to the unperturbed Hamiltonian simply by taking the driving
 amplitude to zero.  This follows from the nature of the identity used to
   derive the Floquet Hamiltonian (see appendix). Here, the
   unkicked case can not be obtained from the case with kicking in the
   straightforward limit of driving amplitude going to zero as the identity
   breaks down in this case. At best a weak driving limit may be assumed
   (which is what the plot indicates) that is physically meaningful. It can be
   seen that substituting $\alpha_z=0$ after evaluating the identity one  can
   not get rid of the periodicity in the phase diagram. Thus the weak driving case essentially is a
   means of, at best, recovering the undriven Haldane model characteristics in
   the presence of kicking. The periodicity will be non-existent in the case
   where there is no kicking  but this
   cannot be approached continuously from the Floquet Hamiltonian which, in
   turn, is not to be regarded as some weak perturbation to the undriven
   case.  One also has to take the limit of the driving period becoming very
 small and ideally going to zero. It is in this limit that one recovers the 
    undriven model's characteristics and this is true for the phases in
 plot (a) of fig.(\ref{fig:fig2}) as the other topological phase regions will
 get pushed out to infinity and one obtains Haldane's original phase
 diagram. An observation that is consistent with the fact that, in the limit
 of driving frequencies being infinitely large, one is precisely left with the
 time-averaged Hamiltonian over a period as the exact description of the
 system. This can be understood in the sense of high frequency expansions
   for effective Hamiltonians where the zeroth order undriven Hamiltonian can
   only be truly retrieved in the large or infinite frequency limit.

 This is so as the
 separation between two pairs of intersecting sinusoids that delineate two topological phase regions is decided by the corresponding driving renormalized 
 Semenoff masses and the difference between these masses can be seen to depend on the driving frequency. Thus one can easily see that the effect of varying
 the driving frequency, say decreasing it in our model, would be to bring the adjacent topological regions, enclosed between their respective pair of sinusoids, nearer to one another.
 Eventually, for the lower driving frequency limit , of which we have spoken earlier, the Haldane-like topological phase diagram copies are close enough for the 
 sinusoids of adjacent diagrams to just touch each other. Going lower in the frequency would take one into the forbidden limit where these non-trivial regions begin  to overlap.

 Further, if one looks at plots (c) and (d) of fig.(\ref{fig:fig2}) we see that for the  driving amplitudes of $\alpha_z=\pi/4$ and $\pi/2$ respectively
 one has  the effect of shifting the topological phases away from the parameter regions which were topologically non-trivial in the undriven situation. Thus in plot
 (c) one can clearly see the new topological region shifted with respect to the phase boundary for the undriven model, which is the pair of sinusoids intersecting
 at the origin in the phase plane. These boundaries are kept specifically to demarcate the topological region of the unkicked Haldane model as reference to
 emphasize the shifting. Even though, as can be seen in the plots, they enclose topologically trivial regions. In particular the upper half of the region enclosed between the undriven model's phase boundaries is now topologically trivial. 
 Thus, increasing the driving, shifts the phases in a linear fashion. One may consider some choice of undriven model parameters $M$ and $\phi$ for which the system is in a topological phase and after a certain magnitude of driving the model enters a topologically trivial phase.
 Thus the change in the driving amplitude can, as discussed earlier, bring a plateau transition in the Chern number. This effect is more pronounced in plot (d), where, the entire parameter range which was topological in the undriven case is now trivial and hence the driving does offer a path to transition  between non-zero and zero Chern numbers and hence may be used to study the normal to Chern insulator transition in such simple non-interacting systems.
  
  Fig.(\ref{fig:fig3}) illustrates the topological phases of the kicked model when looked at from different cross-sectional views of the solid 
  three dimensional structure that would result if the various phase plots for the $\alpha_z$ values , such as those in fig.(\ref{fig:fig2}), were 
  stacked in proper sequence, one above the other, along an out of plane $\alpha_z$ axis. In this figure, all the parameter values that need
  specification to obtain the plots therein, are chosen to be the same as those used for fig.(\ref{fig:fig2}).  Plot (a) in the figure depicts the behavior of the topological
  regions for a $\phi$ value fixed at $\pi/3$ and, $M/t_2$ and $\alpha_z$ being varied. The linear variation of the phases in this picture reveals the 
  linear shift in the sinusoidal lobes seen in the plots of fig.(\ref{fig:fig2}), with change in driving. Additionally, the sharp turn in slope, as if a
  reflection, of these linear phase regions, which are basically tubes with sinusoidal cross-sections, at $\alpha_z=0$ is indicative of the driving dependence 
  being purely on the magnitude i.e. $\vert\alpha_z\vert$. Once this picture is established it becomes easier to interpret the other two plots (b) and (c),
  which show the $\phi-\alpha_z$ phase plane for $M/t_2$ values $2$ and $10$ respectively. Since a constant $M/t_2$ can be understood as a plane that 
  slices a kind of Pan flute structure of the tubes of intersecting sinusoids. Thus on the plane one expects to get the projections of the tubes that are
  cut and this naturally depends on where one chooses to slice. So, where such flutes of different inclination meet, which is at the turning point i.e. $\alpha_z=0$ 
  or, if one considers the full periodicity, $n\pi$, they form an intersecting sinusoidal edge. If the slice is chosen that it cuts above or below the exact
  centre of this ridge i.e. $M\neq0$ then the projection on the corresponding $\phi-\alpha_z$ plane will have a pair of non-touching sinusoids at the centre.
  This is what shows up in the middle of plot (b). Of course the slice may be so chosen that it lies outside this intersecting sinusoidal edge in which case it 
  will cut the nearest sloping flute tubes and result in a projection with touching sinusoids, as is the case in plot (c). Due to the inherent periodicity
  in the phase diagram structure, as one goes through a complete period of the $M/t_2$ choices, the projections begin to show the underlying periodicity.
 
  Before proceeding further, we would like to compare and contrast our
   model with other similar studies in the literature
   \cite{WangLiKickChernPRB17}. In this work the authors look at the impact
   that periodic kicking has on the QWZ Chern insulator, a model similar to
   the Haldane model in that it admits complex valued hoppings, which features
   in the models for the quantum spin Hall effect. The study shows that for
   $\hat{z}$-kicking the QWZ insulator exhibits topological phases with higher
   order Chern numbers. The authors attribute this to merging and splitting of
   Dirac cones and the appearance of long-range hoppings due to driving.  Our
   model, on the other hand, does not exhibit higher order Chern numbers. The
   criteria for band touchings essentially remains the same both before and
   after driving at least to the extent that the number of Dirac points within
   the first Brillouin zone remains unchanged. The Chern numbers associated
   with the topological phases remain the same after driving. Even if long-range
   hoppings do appear, the Chern numbers of the $\hat{z}$-kicked Haldane model
   are robust to these, unlike in the QWZ model. 
 \subsection{Modifications to Haldane's overlap criterion due to kicking}  
 
 This broadly concludes our discussion of the topological features  of the kicked Haldane model. We now turn our attention to the issue of avoiding band overlap in 
 the presence of driving, a concern which has been repeatedly expressed at various points in the above discussion under different contexts. The prime 
 consideration is to have the bands touch in a way that the spectrum allows these touchings to be detected without ambiguity. This imposes a relation on the 
 hopping parameters. As the relative magnitude of $t_1$ and $t_2$ has the effect of  influencing the degree of particle-hole symmetry breaking in the system, and
 hence, the nature of the touchings. Along lines similar to the arguments for Haldane's criterion we obtain the following condition that needs to be satisfied

 \begin{widetext}
 \begin{equation}
 \begin{split}
9t_2 < \cos^{-1}\biggl[\cos(\alpha_z)\cos\left(T\sqrt{9t_1^2+M^2}\right)
  -&\frac{M\sin(\alpha_z)\sin(T\sqrt{9t_1^2+M^2})}{\sqrt{9t_1^2+M^2}}\biggr]\\
  &- \cos^{-1}\biggl[\cos(\alpha_z)\cos(T\vert M\vert) - {\rm sgn}(M)\sin(\alpha_z)\sin(T\vert M\vert)\biggr]
\end{split}
\end{equation}
\end{widetext}

 In the above inequality, a condition is imposed on the suitable values for $t_2$ once $t_1$ has been chosen. This is accompanied with
 the effects of driving also having a role to play in the determination of this value. Both the driving amplitude $\alpha_z$ and the driving period 
 $T$ appear in the above expression. Since we have already done so in our earlier analysis $T$ can be taken to  depend in an appropriate way 
on the nearest neighbor hoppings $t_1$. The $M$ can be written in terms of the driving amplitude using the previously derived expressions for 
the new Semenoff masses $M'$ depending on which $n$-th order semi-metallic center one is looking at to observe the band touchings, by putting 
that particular choice of $M'$ to zero or $n\pi$. Thus the condition can be reduced to depend solely on $\alpha_z$ and $t_1$. Another feature of this condition 
is that unlike the ordinary one given by Haldane which has a simpler reciprocal relationship between the two hopping energies, the above relation is not 
easily invertible to a case where one fixes $t_1$ and calculates the condition on $t_2$. In the context of varying $\alpha_z$ for a fixed $n$ in the 
choice of $M'$ or changing $n$ for fixed $\alpha_z$ the variation in the choice of $t_2$ will have the effect of altering the boundary sinusoids of the
corresponding phase diagrams in the parameter space. Thus if one were to rigorously enforce this condition, which we have ignored for now in the phase diagrams in 
fig.(\ref{fig:fig2}) where $t_2$ is fixed at unity, we would observe a flattening or broadening of the pair of intersecting sinusoids. This follows from
the fact that changing $t_2$ say in the diagram of a given $\alpha_z$ for different $M$ and hence $n$ values would rescale the vertical axis of the diagram.
We would like to point out that adjusting $t_2$ is a freedom available only in certain realizations, as mentioned earlier, hence, if one is interested in 
driving the system across a topological transition it would be reasonable to do so in the previously suggested range of $\alpha_z\in[-1,1]$. Since within
this domain the ordinary haldane condition is a workable choice and one need not be concerned too much about the effects of driving in this regard.

\section{Experimental Aspects}
\label{exp}
 We elaborate here, some of the ideas that were mentioned in the
  introduction relating to the experimental realization of our kicked
  model. The exact realization of a kicking scheme is an experimentally
  formidable challenge owing to finite response and relaxation times of
  physical systems. There are however, a few proposals, that deserve mention
  as they may simulate a periodic driving which can in a suitable
  limit yeild the effects of kicking. From an actual material realization
  perspective, the Silicene based Haldane model of \cite{WrightNatSRep13} can
  be driven using the schemes cited in \cite{AgarwalaDiptimanPRB16} to
  simulate similar kicking but in graphene. In \cite{AgarwalaDiptimanPRB16} the authors cite a method
  that involves placing a sample of hexagonal boron nitride over a layer of
  graphene in such a way that adjacent boron and nitrogen atoms are positioned
  over neighbouring carbon atoms of graphene
  \cite{ShafiqAdamNat15,OrtixYangPRB12,WeinbergStaarmann2DMat16}. Such a
  structural arrangement is permitted by the almost identical lattice
  constants of both materials. This proposal of \cite{AgarwalaDiptimanPRB16}
  is motivated by its capacity to realize an effective sublattice potential
  for the underlying graphene layer. Once this configuration is established
  the driving is implemented by applying periodically varying pressure on the
  two layers such that the distance between them varies periodically in
  time. This would effectively vary the sublattice potential experienced by
  the carbon atoms belonging to the A and B sublattices of graphene and thus
  produce the effects of z-kicking. This could be replicated in the Silicene
  based realization by an appropriate choice of a transition metal
  dichalcogenide that fits the lattice spacing.
  
 The other possible platform for realizing our model is using a cold-atom
 optical lattice setup. Such a quantum simulation of the Haldane model
 \cite{coldatomHaldane} may be subjected to a pulsing that can be designed to
 impart the effects of driving using a periodic train of delta-function kicks
 \cite{SpielmanKickarxiv17}. The technique in its present form simulates a 2-D
 rectangular lattice with a synthetic magnetic field by subjecting a gas of
 ultra-cold alkali metal atoms to a pair of counter-propogating Raman lasers
 in one direction and a magnetic field gradient perpendicular to this. The
 Raman lasers are tuned near resonance in such a way as to give a
 multi-frequency coupling term which, along with the position dependent
 detuning gradient, in the time domain gives a $2\times2$ delta-function
 kicking with a position dependent strength. Though this method cites the lack
 of an initial confining optical lattice as a convenience since this
 eliminates any restrictions on the modulation frequency to avoid transitions
 between the pre-existing Bloch bands, any scheme for realizing our model
 would possibly require the presence of a hexagonal or brick-like optical
 lattice \cite{coldatomHaldane} on which the kicking is realized. Further the
 advantages offered with respect to heating effects due to interactions or
 micromotion could be adversely affected under such modifications.
 
 As regards the possibilities of heating and thermalization in our system,
 certain remarks are in order. The Haldane model being driven is a free
 fermion model in the non-interacting, single particle approximation. The
 overall dynamics of the Floquet system is integrable and hence one in general
 does not expect heating to infinite temperatures over long times, due to the
 driving \cite{GritsevPolkoIntegFloq17}. There is however recent work that
 suggests the possibility of heating in periodically driven systems that are
 integrable \cite{HeatingIntegSys17} owing to divergences in the
 Floquet-Magnus expansion for the effective Hamiltonian at small enough
 driving frequency.  This, however, is not an issue in our closed form
 stroboscopic Floquet Hamiltonian for the kicked system, where although the
 component Hamiltonians describing a single period of evolution do not commute
 with each other the expansion still converges for all frequencies. This
 follows from the identity discussed in the appendix. What has been said thus
 far takes the viewpoint of the formalism that is used to describe our system
 where, in the absence of interactions or disorder, there is little likelihood
 of many-body localized phases. Though one may reasonably expect to find
 dynamically localized wavefunctions as was demonstrated in
 \cite{AgarwalaDiptimanPRB16}. But from an experimental viewpoint, say in a
 material realization, these assumptions may not hold and heating could
 occur. In such situations choosing the driving frequency to be high would
 help to make the heating rate exponentially slow, as it decays exponentially
 in the driving frequency. The heating time is therefore large and one should
 be able to observe the Floquet topological phenomena on experimentally
 resonable time scales that are not exponentially large in the frequency. In
 this case the existence of many-body localized phases is a distinct
 possibility especially at large frequencies, ensuring the system does not
 thermalize. Though this might require certain restrictions on system size and
 hopping/interaction energies when one speaks of finite size
 systems. Although, in our formalism, we work with the assumption of infinite
 system size, as can be seen from the hamiltonian for the Haldane model. The
 authors in \cite{HeatingIntegSys17} suggest heating to infinite temperatures
 for Floquet systems with integrability as an asymptotic phenomeneon in the
 limit of driving period and system size going to infinity. This too is
 observed only for choices of random or quasiperiodic fields within a single
 driving period. For the case of a staggered field the long time steady state
 is never found to correspond to an infinite temperature state even for the
 infinite frequency and size limits. This suggests that even our kicked
 infinite system should be immune to the possibility of heating to infinite
 temperature, from a theoretical perspective.

\section{Conclusion}
 We have considered  a $\hat{z}$-kicked Haldane model and examine the topological properties of this system. The effects of driving on the topological phase diagram
 of Haldane's originally proposed model are illustrated. We find that, besides introducing a periodicity in the phase diagram where the Haldane phase diagram 
 is repeated at regular intervals along the inversion breaking axis $M/t_2$, a signature of the periodicity of the Floquet quasienergy spectrum, the driving
 magnitude is solely responsible for a linear shift in the topological phases of the driven model relative to their undriven counterparts. This suggests the use
 of this driven model to study Floquet topological phase transitions.  
 
 This is different from the optically driven Haldane models of \cite{InoueTanakaPRL10,WangLiWuEPL12} where the tunable parameter toys with the time-reversal symmetry
breaking by modifying those terms of the effective Hamiltonian which depend on the phase of the complex valued next-nearest neighbor hoppings of the undriven Haldane
model. Although the overall effect is to still traverse between topological and non-topological regions of the Haldane Chern number phase plot as drawn against the
symmetry breaking parameters, yet this is brought about in a different manner. To be precise this distinction becomes fully apparent when one considers the effective 
Hamiltonian post-driving in the vicinity of a Dirac point which of course would be usually gapped in the given case. The manner in which the topological phases transition with change in
driving amplitude is also different in \cite{InoueTanakaPRL10}, compared to our case, as there, the sinusoids of the topological phase boundaries undergo a contraction-expansion
in a periodic manner instead of the linear shifting that is seen in our case. Also, the driving at sufficiently 
large amplitudes causes a modification of the band overlap avoidance criterion as originally suggested by Haldane in his model. 

Finally, we would like to mention that this kicking scheme could be also applied to the Kane-Mele model for spin orbit coupling in hexagonal lattices \cite{KaneMelePRL05,KaneMeleZ2PRL05} to study
the effect of driving on the $Z_2$ topological index which characterizes the topology in such QSHE systems. This is proposed as a future work that we intend to
undertake. 

\section{Acknowledgements}
T.M. thanks UGC, India for funding through a SRF and would like to acknowledge discussions with Prof. Diptiman Sen, Prof. Amit Dutta and Dr. Utso Bhattacharya in meetings at ICTS. A.P. would like to thank
Dr. Adhip Agarwala for valuable inputs. T.G.S and J.N.B thank DST-SERB, India for Project No. EMR/2016/003289. They would also like to thank Prof. T. Oka for discussions and Prof. N. Mukunda for providing his lecture notes on  various aspects of geometric phase.
 \appendix*
\section{}
\label{append}
 In order to obtain the Floquet Hamiltonian for the kicked system we make use of a well known identity from the algebra of Pauli matrices, which allows
one to write 
\begin{equation}
\label{ident1}
 e^{ip(\hat{n}\cdot\boldsymbol{\sigma})}e^{iq(\hat{m}\cdot\boldsymbol{\sigma})}= e^{ig(\hat{l}\cdot\boldsymbol{\sigma})}.
\end{equation}
Where, $p,q$ and $g$ are scalars and $\hat{n},\hat{m}$ and $\hat{l}$ are unit vectors and for known $p,q$ and $\hat{n},\hat{m}$, $g$ and $\hat{l}$ are given 
by the relations 
\begin{align}
\label{ident2}
 \cos g &= \cos p\cos q - \hat{n}\cdot\hat{m}\sin p \sin q \nonumber\\
 \hat{l}&=  \frac{1}{\sin g}(\hat{n}\sin p\sin q +\hat{m}\sin q\cos p\nonumber\\
 &-\hat{n}\times\hat{m}\sin p\sin q)
\end{align}

From section \ref{Halmod} we see that the time evolution over a period is  $U_{XYZ} = U_{kick}U_{static} = e^{-i\boldsymbol{\alpha}.\boldsymbol{\sigma}}e^{-i\mathcal{H}(\mathbf{k})T}$.
We shall use the above identity to compute the Floquet Hamiltonian $\mathcal{H}_{XYZ}$, such that $U_{XYZ} = e^{-i\mathcal{H}_{XYZ}(\mathbf{k})T}$. To do this we set up
the correspondence between the quantities of the evolution operator and the variables in the LHS of eq.\eqref{ident1} as, 
\begin{align}
 p &= -\vert\boldsymbol{\alpha}\vert\nonumber\\
 q &= -T\vert \mathbf{h}(\mathbf{k})\vert\nonumber\\
\hat{n} &= \frac{\boldsymbol{\alpha}}{\vert\boldsymbol{\alpha}\vert}\nonumber\\
\hat{m} &= \frac{1}{\vert \mathbf{h}(\mathbf{k})\vert}(t_1L(\mathbf{k}),t_1F(\mathbf{k}), M-2t_2N(\mathbf{k})\sin\phi )
\end{align}
On substituting the above in the relations of eq.\eqref{ident2} we can compute the values of $g$ and $\hat{l}$ respectively. We are interested in the case of
$\hat{z}$-kicking and so one needs to additionally impose the conditions $\alpha_z\neq0$ while $\alpha_x=\alpha_y=0$. The Floquet Hamiltonian for this case 
is denoted by $\mathcal{H}_{Z}$. In our case $g$ corresponds to the Floquet spectrum as $g= T\epsilon_z(\mathbf{k})$ and the unit vector $\hat{l}$ to $\mathbf{h'}(\mathbf{k})$, as can be seen from 
section \ref{Halmod}. Thus we get the vector components $(h'_x(\mathbf{k}),h'_y(\mathbf{k}),h'_z(\mathbf{k}))$  in eq.\eqref{KickedHalcomp} and the Floquet band
spectrum $\epsilon_z(\mathbf{k})$ in eq.\eqref{Floqspec}.
 \bibliographystyle{apsrev4-1}
 \bibliography{Reference}

\end{document}